\documentclass[11pt,a4paper]{article}

\usepackage{amsmath,amssymb,epsfig,a4,latexsym,axodraw}
\usepackage{mathtools}
\usepackage{paralist}
\usepackage{slashed}
\usepackage{slantsc}
\usepackage{jheppub}
\usepackage{subfigure}
\usepackage{etoolbox}
\usepackage{array}
\usepackage{enumitem}
\usepackage{pifont}

\allowdisplaybreaks[1]
\newcommand{\defeq}{\mathrel{\mathop:}=}

\def\HV{{\scshape hv}}
\def\FDH{{\scshape fdh}}
\def\DRED{{\scshape dred}}
\def\FDR{{\scshape fdr}}

\def\Ahat{{\hat{A}}}
\def\Ahatslashed{{\mathrlap{\not{\phantom{A}}}\hat{A}}}
\def\Atilde{{\tilde{A}}}
\def\Atildeslashed{{\mathrlap{\not{\phantom{A}}}\tilde{A}}}
\def\Dhat{{\hat{D}}}

\def\Fhat{{\hat{F}}}

\def\Otilde{{\tilde{O}}}

\newcommand{\RS}{\overline{\text{MS}}}

\newcommand{\Neps}{N_\epsilon}
\newcommand{\alphas}{\alpha_s}
\newcommand{\alphae}{\alpha_e}
\newcommand{\alphafour}{\alpha_{4\epsilon}}
\newcommand{\HiggsGlu}{\lambda}
\newcommand{\HiggsEps}{\lambda_{\epsilon}}
\newcommand{\alphaFourEps}{\alpha_{4\epsilon}}

\def\gs{g_{s}}
\def\ge{g_{e}}
\def\gfour{g_{4\epsilon}}
\def\lambdaeps{\lambda_{\epsilon}}
\def\leps{\lambda_{\epsilon}}
\def\gammahat{\hat{\gamma}}

\def\ghat{{\hat{g}}}
\def\gtilde{{\widetilde{g}}}

\def\gbar{{\bar{g}}}

\def\pslash#1{{\setbox0=\hbox{$#1$}
  \rlap{\ifdim\wd0>.7em\kern.22\wd0\else\kern.1\wd0\fi /}#1}}

\def\ghat{{\hat{g}}}
\def\gtilde{{\tilde{g}}}
\def\gbar{{\bar{g}}}

\def\DRED{{\scshape dred}}

\def\HV{{\scshape hv}}
\def\FDH{{\scshape fdh}}
\def\CDR{{\scshape cdr}}

\def\LeffOp{{\cal L}_{\text{eff}}}

\DeclareMathAlphabet{\mathbbmsl}{U}{bbm}{m}{sl}

\usepackage{datetime}
\begin{document}
\thispagestyle{empty}
\begin{flushright}
PSI-PR-15-02 \\
ZU-TH 05/15 \\
\end{flushright}
\vspace{3em}
\begin{center}
{\Large\bf Computation of $H\to gg$ in FDH and DRED:
  renormalization, operator mixing, and explicit two-loop results}
\\
\vspace{3em}
%{\large {\renewcommand{\thefootnote}{\fnsymbol{footnote}}
%\footnote{\parbox[t]{10cm}{}}}}
%  \\[2ex]
%  \parbox{10cm}{%\small\center\em
{\sc A.\ Broggio$^a$, Ch.\ Gnendiger$^b$, A.\ Signer$^{a,c}$, D. St\"ockinger$^b$, A.\ Visconti$^a$
}\\[2em]
{\sl
${}^a$ Paul Scherrer Institut,\\
CH-5232 Villigen PSI, Switzerland \\
\vspace{0.3cm}
${}^b$ Institut f\"ur Kern- und Teilchenphysik,\\
TU Dresden, D-01062 Dresden, Germany\\
\vspace{0.3cm}
${}^c$ Physik-Institut, Universit\"at Z\"urich, \\
Winterthurerstrasse 190,
CH-8057 Z\"urich, Switzerland}
\setcounter{footnote}{0}
\end{center}
\vspace{2ex}
\begin{abstract}
{}The $H\to gg$ amplitude relevant for Higgs production via gluon fusion
is computed  in the four-dimensional helicity scheme (\FDH) and
in dimensional reduction (\DRED) at the two-loop level.
The required renormalization is developed and
described in detail, including the treatment of evanescent
$\epsilon$-scalar contributions.
In \FDH\ and \DRED\ there are additional dimension-5 operators
generating the $H g g$ vertices, where $g$ can either be a gluon or an
$\epsilon$-scalar. An appropriate operator basis is given and the
operator mixing through renormalization is described.
%required mixing renormalization constants are given.
The results of the present paper provide building blocks for further
computations, and they allow to complete the study of the infrared
divergence structure of two-loop amplitudes in \FDH\ and \DRED.
\end{abstract}

\newpage
\noindent\hrulefill
\tableofcontents
\noindent\hrulefill

\section{Introduction}

Higgs production via gluon fusion is one of the most
important LHC processes. Its computation at higher orders requires
renormalization and factorization to cancel UV and IR divergences. The
renormalization is less trivial than the one of standard QCD processes
due to the required renormalization of non-renormalizable
operators. The virtual corrections have been computed in conventional dimensional
regularization (\CDR)~\cite{Harlander:2000mg,Moch:2005tm,Baikov:2009bg,
Gehrmann:2010ue,Gehrmann:2010tu};
the required theory of operator renormalization in \CDR\ has been developed
in Ref.~\cite{Spiridonov:1984br}, based on general work in
Refs.~\cite{KlubergStern:1974rs,Joglekar:1975nu}. 

In the past years, several alternative regularization schemes have
been developed. Purely four-dimensional schemes such as implicit
regularization \cite{Cherchiglia:2010yd,Ferreira:2011cv} and
\FDR~\cite{Pittau:2012zd} have been proposed and
used to compute processes of practical interest such as
$H\to\gamma\gamma$ \cite{Cherchiglia:2012zp,Donati:2013iya} and $H\to gg$
\cite{Pittau:2013qla}.
The present paper is devoted to regularization
by dimensional reduction (\DRED) \cite{Siegel:1979wq}
and the related four-dimensional helicity (\FDH) scheme \cite{BernZviKosower:1992}.
Both schemes are actually the same regarding UV renormalization,
but they differ in the treatment of external partons related to IR
divergences.\footnote{Parts of the literature, e.\,g.
Refs.~\cite{Kunszt:1993sd, Catani:2000ef, Catani:1996pk}
used the term DR/dimensional reduction for what is called \FDH\ here.
}
There has been significant progress
in the understanding of \FDH\ and \DRED:
the equivalence to \CDR\ \cite{Jack:1993ws,Jack:1994bn},
mathematical consistency and the quantum action principle
\cite{Stockinger:2005gx},
infrared factorization \cite{Signer:2005,Signer:2008va} have been established ---
these results solved several problems that had been reported earlier,
related to violation of unitarity \cite{vanDamme:1984ig}, Siegel's
inconsistency \cite{Siegel:1980qs}, and the factorization problem of
\cite{Beenakker:1989, Smith:2005}. In addition, explicit multi-loop calculations
have been carried out \cite{Harlander:2006rj,Harlander:2006xq,Kant:2010tf,
Kilgore:2011ta,Kilgore:2012tb}.

More recently, the multi-loop IR divergence structure of \FDH\ and \DRED\
amplitudes has been studied in Ref.~\cite{Gnendiger:2014nxa}.
It has been shown that IR divergences in \FDH\ and \DRED\ can be
described by a generalization of the \CDR\ formulas given in
Refs.~\cite{Becher:2009qa,Becher:2009cu,Gardi:2009zv,Magnea:2012pk}.
The description involves IR anomalous dimensions $\gamma^i$ for each
parton type $i$.
In Ref.~\cite{Gnendiger:2014nxa} they have been computed for the cases
of quarks and gluons by comparing the general IR factorization formulas
with explicit results for the quark and gluon form factor.
In \FDH\ and \DRED, however, the
gluon can be decomposed into a $D$-dimensional gluon $\ghat$ and
$(4-D)$ additional degrees of freedom, so-called $\epsilon$-scalars
$\gtilde$. In \DRED, $\epsilon$-scalars also appear as external states.

The present paper is devoted to a detailed two-loop computation of the
amplitude $H\to gg$ in \FDH\ and \DRED. In \DRED, this involves the
computations of $H\to \ghat\ghat$ and
$H\to\gtilde\gtilde$, since the external gluons can either be
gauge fields or $\epsilon$-scalars. The \FDH\ result is identical to
the one for $H\to \ghat\ghat$ and has already been given in
Ref.~\cite{Gnendiger:2014nxa}, but we will provide further details
here.

This detailed computation is of interest for two
reasons: First, it provides the basis for obtaining the remaining IR
anomalous dimension for $\epsilon$-scalars at the two-loop
level. Second, it provides an example of the required renormalization
in \FDH\ and \DRED, including operator renormalization and operator
mixing. The difficulty of renormalization in \FDH\ and \DRED,
particularly in connection with $H\to gg$, has been pointed out
e.\,g. in Refs.~\cite{Kilgore:2012tb, Anastasiou:2008rm}.

The outline of the paper is as follows: Section \ref{sec:setup} gives a brief
description of the regularization schemes and of the
relevant Lagrangian and operators.
It ends with a detailed list
of the required ingredients of the calculation.
% and an outline of
%Sections \ref{sec:two-loop}--\ref{sec:CT2lb}, which
%present those ingredients in detail.
Apart from the actual two-loop computation and ordinary parameter and field
renormalization that are described in Sections \ref{sec:two-loop} and 
\ref{sec:fieldparameterrenormalization}, respectively,
the main difficulty lies in the renormalization and
mixing of the operators generating $H\to gg$. This is discussed in
general in Section \ref{sec:operatorrenormalization},
and specific two-loop results are presented in
Section \ref{sec:CT2Lb}.
Section \ref{sec:onshell_results} then provides the final results for the on-shell
amplitudes for $H\to \ghat\ghat$ and $H\to\gtilde\gtilde$.
The appendix contains details on our projection operators and gives Feynman
rules for the different operator insertions.

\section{Regularization schemes and $H\to gg$}
\label{sec:setup}

It is useful to distinguish the following regularization schemes
\cite{Signer:2008va}:
conventional dimensional regularization (\CDR),
the 't Hooft-Veltman (\HV) scheme,
the four-dimensional helicity (\FDH) scheme, and
dimensional reduction (\DRED). In all these schemes, momenta are treated in
$D=4-2\epsilon$ dimensions (the associated space is denoted by $QDS$ with
metric tensor 
$\ghat^{\mu\nu}$). In order to define the schemes, one also needs an
additional quasi-4-dimensional space ($Q4S$, metric $g^{\mu\nu}$) and the
original 4-dimensional space ($4S$, metric $\gbar^{\mu\nu}$).
The treatment of gluons in the four schemes is given in Tab.~\ref{tab:RSs}. 
In the table,
``internal'' gluons are defined as either virtual gluons that are part of a
one-particle irreducible loop diagram or, for real correction
diagrams, gluons in the initial or final state that are collinear or
soft. ``External gluons'' are defined as all other gluons. 
\begin{table}
\begin{center}
\begin{tabular}{l|cccc}
&\CDR&\HV&\FDH&\DRED\\
\hline
internal gluon&$\ghat^{\mu\nu}$&$\ghat^{\mu\nu}$&
$g^{\mu\nu}$&$g^{\mu\nu}$\\
 external gluon&$\ghat^{\mu\nu}$&$\gbar^{\mu\nu}$&
$\gbar^{\mu\nu}$&$g^{\mu\nu}$
\end{tabular}
\end{center}
\caption{
Treatment of internal and external gluons in the four different
regularization schemes,
i.e.\ prescription which metric tensor has to be used in propagator
numerators and polarization sums.
%For the definition of ``internal'' and ``external'' see the text.
\label{tab:RSs}
}
\end{table}

Mathematical consistency and $D$-dimensional gauge invariance require
that $Q4S\supset QDS\supset 4S$ and  forbid to identify
$g^{\mu\nu}$ and $\gbar^{\mu\nu}$. Details can be found
in Refs.\ \cite{Stockinger:2005gx,Signer:2008va,Gnendiger:2014nxa}.
The most important relations for the present paper are
\begin{align}
g^{\mu\nu}&=\ghat^{\mu\nu}+\gtilde^{\mu\nu},&
\ghat^{\mu\rho}\gtilde_{\rho}{}^\nu&=0,&
\ghat^{\mu\rho}\bar{g}_{\rho}{}^\nu&=\bar{g}^{\mu\nu},&
\ghat^{\mu\nu}\ghat_{\mu\nu}&=D,&
\gtilde^{\mu\nu}\gtilde_{\mu\nu}&=\Neps,&
\end{align}
where a complementary $2\epsilon$-dimensional metric tensor
$\gtilde^{\mu\nu}$ has been introduced. With these metric tensors we
can decompose a quasi-4-dimensional gluon field $A^\mu$ as
\begin{align}
A^\mu&=\ghat^{\mu\nu}A_\nu+\gtilde^{\mu\nu}A_\nu=\hat{A}^\mu+\tilde{A}^\mu
\end{align} 
into  a $D$-dimensional gauge field $\hat{A}^\mu$ and an associated
$\epsilon$-scalar field  $\tilde{A}^\mu$ with
multiplicity~$\Neps=2\epsilon$.\,\footnote{%
In many applications of \FDH\ the dimensionality of $Q4S$ is left as a
variable $D_s$, which is eventually set to $D_s=4$. The multiplicity
of $\epsilon$-scalars is then $\Neps=D_s-D$.}
Correspondingly, there are two types of
particles in the regularized theory: $D$-dimensional gluons $\ghat$
and $\epsilon$-scalars $\gtilde$. The unregularized external gluons
$\bar{g}$ of \FDH\ are a part of $\ghat$.

The regularized Lagrangian of massless QCD then reads
\begin{subequations}
\begin{align}
 {\cal{L}}_{QCD,\text{ regularized}} &= -\frac{1}{4}\Fhat^{\mu\nu}_a\Fhat_{\mu\nu,a}
   -\frac{1}{2\xi}(\partial^\mu\Ahat_{\mu,a})^2
   +i\,\overline{\psi}\hat{\slashed{D}}\psi
   +\partial^\mu\overline{c}_{a}\Dhat_\mu c_{a}+\cal{L}_\epsilon,\\[15pt]
 \cal{L}_\epsilon &=
   -\frac{1}{2}(\hat{D}^\mu\Atilde^\nu)_a(\hat{D}_\mu\Atilde_\nu)_{a}
   -\ge\,\overline{\psi}\Atildeslashed\psi
   -\frac{1}{4!}\left(\gfour^2\right)^{\alpha\beta\gamma\delta}_{abcd}
     \Atilde_{\alpha,a}\Atilde_{\beta ,b}
     \Atilde_{\gamma,c}\Atilde_{\delta,d}.
\label{LQCDregularized}
\end{align}
\end{subequations}
Here, $\Fhat^{\mu\nu}$ and $\Dhat^\mu=\partial^\mu+i\gs\Ahat^\mu$ denote the
non-abelian field strength tensor and the covariant derivative in $D$ dimensions;
$\psi$ and $c$ are the quark and ghost fields.
In Eq.~(\ref{LQCDregularized}) the coupling of $\epsilon$-scalars to (anti-)quarks is given by
the evanescent Yukawa-like coupling~$\ge$. This could in principle be
set equal to the strong coupling~$\gs$. But, since both couplings renormalize differently
this would only hold at tree-level and for one particular renormalization 
scale~\cite{Jack:1993ws};
the same is true for the quartic $\epsilon$-scalar coupling $\gfour$.
In Eq.~(\ref{LQCDregularized}) we introduce an abbreviation that includes the appearing
Lorentz and color structure:
$\left(\gfour^2\right)^{\alpha\beta\gamma\delta}_{abcd}\defeq
\gfour^2(f_{abe}f_{cde}
\gtilde^{\alpha\gamma}\gtilde^{\beta\delta}+\text{perm.})$, where ``perm.''
denotes the 5 permutations arising from symmetrization in the multi-indices
$(a,\alpha)\dots(c,\gamma)$.
In the following we use all couplings in the form
$\alpha_i=\frac{g_i^2}{4\pi}$ with $i = s, e, 4\epsilon$.

The process $H\to gg$ is generated by an effective Lagrangian which
arises from integrating out the top quark in the Standard Model. In
\CDR\ it contains only the term
$-\frac{1}{4}\lambda H \Fhat^{\mu\nu}_a\Fhat_{\mu\nu, a}$.
In \FDH\ and \DRED\ one again has to distinguish several gauge invariant
structures containing either $D$-dimensional gluons or
$\epsilon$-scalars. The effective Lagrangian can be written as
\begin{align}
\LeffOp &=
\lambda HO_1+\lambda_\epsilon H\tilde{O}_1+
\sum_i\lambda_{4\epsilon,i}H\tilde{O}_{4\epsilon,i},
\label{Leff}
\end{align}
with
\begin{subequations}
\begin{align}
O_1 &=-\frac{1}{4} \hat{F}^{\mu\nu}_a \hat{F}_{\mu\nu,a},\\
\Otilde_1 &=
  -\frac{1}{2}(\hat{D}^\mu\Atilde^\nu)_a(\hat{D}_\mu\Atilde_\nu)_a.
\end{align}
\end{subequations}
$\Otilde_{4\epsilon,i}$ denote operators involving
products of four $\epsilon$-scalars. %, i.e.\ of ${\cal O}(\tilde{A}^4)$
Such operators are not important in the present paper and will
not be given explicitly.
Like for $\alpha_s,\alphae$ and $\alphaFourEps$, the couplings $\lambda$ and
$\lambda_\epsilon$ can be set equal at tree-level, but they
renormalize differently and have different $\beta$ functions.

Our final goal is the calculation of the two-loop form factors for
gluons and $\epsilon$-scalars. This requires the
on-shell calculation of the 3-point function 
$\Gamma_{H A^\mu A^\nu}(q,-p,-r)$. All momenta are
defined as incoming, so $q=p+r$. The 3-point function can be separated
into $\Gamma_{H \Ahat^\mu \Ahat^\nu}$ and $\Gamma_{H \Atilde^\mu
\Atilde^\nu}$, corresponding to the amplitudes for $H\to \ghat\ghat$ and
$H\to\gtilde\gtilde$, respectively. In \DRED, both on-shell amplitudes
are needed according to Tab.\ \ref{tab:RSs}. In \FDH, only $H\to
\bar{g}\bar{g}$ is needed, which however is identical to
$H\to\ghat\ghat$ and will not be discussed seperately.

The on-shell calculation requires the knowledge of the two-loop
renormalization constants $\delta Z_\lambda^{\text{2L}}$ and
$\delta Z_{\lambda_\epsilon}^{\text{2L}}$.
These in turn can be obtained from an off-shell calculation of
$\Gamma_{H A^\mu A^\nu}$. Projectors extracting the required
renormalization constants from the off-shell Green functions
and precisely defining the gluon and $\epsilon$-scalar form factors are
given in appendix~\ref{sec:appendix_A}.

We have now all ingredients to discuss the classes of Feynman diagrams
that contribute to $\Gamma_{H A^\mu A^\nu}$ in \FDH\ and \DRED:

\begin{enumerate}
\item Genuine two-loop diagrams $\Gamma_{H A^\mu A^\nu}^{\text{2L}}$.
  Some remarks concerning the calculation are presented in 
  Sec.~\ref{sec:two-loop}.
\item Counterterm diagrams $\Gamma_{H A^\mu A^\nu}^{\text{1LCT,a}}$
  and $\Gamma_{H A^\mu A^\nu}^{\text{2LCT,a}}$
  arising from one- and two-loop renormalization
  of the fields, the gauge parameter $\xi$, and of the couplings
  $\alpha_s$, $\alphae$, and $\alphaFourEps$.
  The required renormalization constants are
  presented in Sec.~\ref{sec:fieldparameterrenormalization}.
\item Counterterm diagrams $\Gamma_{H A^\mu A^\nu}^{\text{1LCT,b}}$
  arising from one-loop renormalization of the effective Lagrangian
  (\ref{Leff}) 
  at the one-loop level, which includes the renormalization of
  $\lambda$ and $\leps$.  This is a major complication and will be presented in
  Sec.~\ref{sec:operatorrenormalization}. 
\item Overall two-loop counterterm diagrams $\Gamma_{H A^\mu A^\nu}^{\text{2LCT,b}}$
  arising from the two-loop renormalization of the effective
  Lagrangian (\ref{Leff}), equivalently from the 
  renormalization constants
  $\delta Z_{\lambda}^{\text{2L}}$ and $\delta Z_{\lambda_\epsilon}^{\text{2L}}$.
  These  renormalization constants are generally defined by the requirement
  that the appropriate off-shell Green functions are UV finite after
  renormalization. For the case of $\delta Z_\lambda$, an elegant
  alternative determination is possible~\cite{Spiridonov:1984br}, but that
  method fails for $\delta Z_{\lambda_\epsilon}$. The results for
  $\delta Z_{\lambda}^{\text{2L}}$ and
  $\delta Z_{\lambda_\epsilon}^{\text{2L}}$ are
  presented in Sec.~\ref{sec:CT2Lb}.
\end{enumerate}

\section{Genuine two-loop diagrams}
\label{sec:two-loop}

\begin{figure}[t]
\begin{center}
\scalebox{.70}{
\begin{picture}(135,90)(0, 0)
\Vertex(25,45){2}
\Vertex(62.5,62.5){2}
\Vertex(62.5,27.5){2}
\Vertex(100,80){2}
\Vertex(100,10){2}
\DashLine(0,45)(25,45){2}
\Gluon(25,45)(62.5,62.5){4}{3}
\Gluon(25,45)(62.5,27.5){4}{3}
\ArrowLine(100,10)(100,80)
\ArrowLine(100,80)(62.5,62.5)
\ArrowLine(62.5,62.5)(62.5,27.5)
\ArrowLine(62.5,27.5)(100,10)
\Gluon(100,80)(130,80){4}{3}
\Gluon(100,10)(130,10){4}{3}
\Text(10,50)[b]{\scalebox{1.42}{$H$}}
\end{picture}
\quad
\begin{picture}(135,90)(0, 0)
\Vertex(25,45){2}
\Vertex(62.5,62.5){2}
\Vertex(62.5,27.5){2}
\Vertex(100,80){2}
\Vertex(100,10){2}
\DashLine(0,45)(25,45){2}
\DashLine(62.5,62.5)(25,45){4}
\DashLine(25,45)(62.5,27.5){4}
\ArrowLine(100,10)(100,80)
\ArrowLine(100,80)(62.5,62.5)
\ArrowLine(62.5,62.5)(62.5,27.5)
\ArrowLine(62.5,27.5)(100,10)
\DashLine(100,80)(130,80){4}
\DashLine(130,10)(100,10){4}
\Text(10,50)[b]{\scalebox{1.42}{$H$}}
\end{picture}
\quad
\begin{picture}(135,90)(0,0)
\Vertex(30,45){2}
\Vertex(60,45){2}
\Vertex(90,45){2}
\DashCArc(45,45)(15,0,180){3}
\DashCArc(45,45)(15,180,360){3}
\GlueArc(75,45)(15,0,180){4}{4}
\GlueArc(75,45)(15,180,360){4}{4}
\DashLine(0,45)(30,45){2}
\Gluon(90,45)(130,70){4}{4}
\Gluon(90,45)(130,20){4}{4}
\Text(10,50)[b]{\scalebox{1.42}{$H$}}
\end{picture}
\quad
\begin{picture}(135,90)(0, 0)
\Vertex(30,45){2}
\Vertex(60,45){2}
\Vertex(90,45){2}
\DashCArc(45,45)(15,0,180){3}
\DashCArc(45,45)(15,180,360){3}
\DashCArc(75,45)(15,0,180){3}
\DashCArc(75,45)(15,180,360){3}
\DashLine(0,45)(30,45){2}
\DashLine(90,45)(130,70){4}
\DashLine(90,45)(130,20){4}
\Text(10,50)[b]{\scalebox{1.42}{$H$}}
\end{picture}
}
%\end{tabular}
\caption{\label{fig:twoloopdiagrams}
Sample two-loop diagrams for the process $H\to \ghat\ghat$ and
$H\to\gtilde\gtilde$ in \DRED. 
$\epsilon$-scalars are denoted by dashed lines. The appearing coupling
combinations from left to right are $\lambda\alpha_s^2$, $\lambda_\epsilon
\alphae^2$, $\lambda_\epsilon\alpha_s^2$,  $\lambda_\epsilon\alphaFourEps^2$.
}
\end{center}
\end{figure}
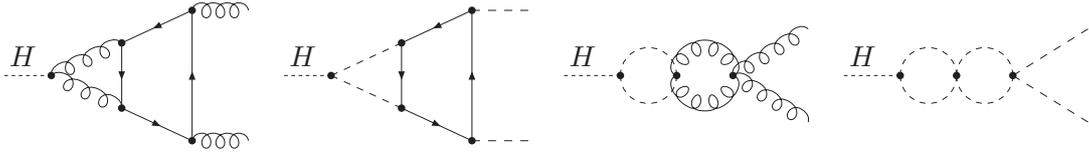

As mentioned above the Green function $\Gamma_{H A^\mu A^\nu}$ can be separated into
$\Gamma_{H \Ahat^\mu \Ahat^\nu}$ and $\Gamma_{H \Atilde^\mu \Atilde^\nu}$,
corresponding to $H\to \ghat\ghat$ and $H\to\gtilde\gtilde$.
Examples for genuine two-loop diagrams with either external gluons or $\epsilon$-scalars
are shown in Fig.~\ref{fig:twoloopdiagrams}.
%They illustrate the different coupling combinations
%$\lambda\alpha_s^2$, $\lambda_\epsilon\alphae^2$, $\lambda_\epsilon\alpha_s^2$, and 
%$\lambda_\epsilon\alphaFourEps^2$, respectively.

%To perform the calculations we used the following setup: the generation of the diagrams
%and the implementation of the Feynman rules is done with the Mathematica package FeynArts~\cite{Hahn:2000kx};
%the subsequent evaluation of the algebra in $D$ and $4$ dimensions is then performed with the package
%TRACER~\cite{Jamin:1991dp}.
%For the reduction and evaluation of the planar integrals we implemented an in-house algorithm
%based on integration-by-parts methods and the Laporta-algorithm~\cite{Laporta:2001dd}.
%The non-planar integrals were reduced and evaluated with the packages
%FIRE~\cite{Smirnov:2008iw} and FIESTA~\cite{Smirnov:2008py}, respectively.

All loop calculations have been performed using the following setup: the generation of diagrams
and analytical expressions is done with the Mathematica package FeynArts~\cite{Hahn:2000kx};
to cope with the extended Lorentz structure in $Q4S$ we use a modified version of TRACER~\cite{Jamin:1991dp};
all planar on-shell integrals are reduced and evaluated with an inplementation of an
in-house algorithm that is based on integration-by-parts methods and the
Laporta-algorithm~\cite{Laporta:2001dd};
all non-planar and off-shell integrals are reduced and evaluated with the packages
FIRE~\cite{Smirnov:2008iw} and FIESTA~\cite{Smirnov:2008py}.

\section{Parameter and field renormalization in FDH and DRED}
\label{sec:fieldparameterrenormalization}

\begin{figure}[t]
\begin{center}
\scalebox{.70}{
\begin{picture}(135,90)(0,0)
\Vertex(25,45){2}
\Text(63,55)[b]{\scalebox{2}{\ding{53}}}
\Vertex(62.5,62.5){2}
\Vertex(62.5,27.5){2}
\DashLine(0,45)(25,45){2}
\Gluon(25,45)(62.5,62.5){4}{4}
\Gluon(25,45)(62.5,27.5){4}{4}
\DashLine(62.5,62.5)(100,80){4}
\DashLine(62.5,27.5)(100,10){4}
\DashLine(62.5,27.5)(62.5,62.5){4}
\Text(10,50)[b]{\scalebox{1.42}{$H$}}
\end{picture}
\quad
\begin{picture}(135,90)(0,0)
\Vertex(25,45){2}
\Text(63,55)[b]{\scalebox{2}{\ding{53}}}
\Vertex(62.5,62.5){2}
\Vertex(62.5,27.5){2}
\DashLine(0,45)(25,45){2}
\ArrowLine(62.5,62.5)(25,45)
\ArrowLine(25,45)(62.5,27.5)
\ArrowLine(62.5,27.5)(62.5,62.5)
\DashLine(62.5,62.5)(100,80){4}
\DashLine(62.5,27.5)(100,10){4}
\Text(10,50)[b]{\scalebox{1.42}{$H$}}
\end{picture}
\quad
\begin{picture}(135,90)(0,0)
\Vertex(30,45){2}
\Vertex(70,45){2}
\Text(70,38)[b]{\scalebox{2}{\ding{53}}}
\DashCArc(50,45)(20,0,180){4}
\DashCArc(50,45)(20,180,360){4}
\DashLine(0,45)(30,45){2}
\DashLine(70,45)(100,10){4}
\DashLine(70,45)(100,90){4}
\Text(10,50)[b]{\scalebox{1.42}{$H$}}
\end{picture}
\quad
\begin{picture}(135,90)(0,0)
\Vertex(25,45){2}
\Text(43,47)[b]{\scalebox{2}{\ding{53}}}
\Vertex(62.5,62.5){2}
\Vertex(62.5,27.5){2}
\DashLine(0,45)(25,45){2}
\Gluon(25,45)(62.5,62.5){4}{4}
\Gluon(25,45)(62.5,27.5){4}{4}
\DashLine(62.5,62.5)(100,80){4}
\DashLine(62.5,27.5)(100,10){4}
\DashLine(62.5,27.5)(62.5,62.5){4}
\Text(10,50)[b]{\scalebox{1.42}{$H$}}
\end{picture}
}
\caption{\label{fig:CTa}
Sample one-loop counterterm diagrams originating from the renormalization
of the couplings $\alphas, \alphae$, $\alphafour$, and of the gauge
parameter $\xi$, respectively.
}
\end{center}
\end{figure}
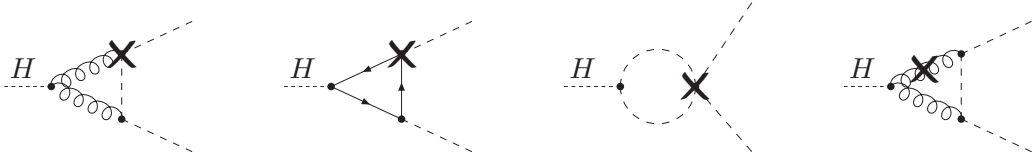

We now consider the counterterm contributions
$\Gamma_{H A^\mu A^\nu}^{\text{1LCT,a}}$
and $\Gamma_{H A^\mu A^\nu}^{\text{2LCT,a}}$. They are given by
diagrams exemplified in Fig.~\ref{fig:CTa}, where the counterterm
insertions are generated by the usual multiplicative QCD renormalization
of the couplings and fields present in Eq.~(\ref{LQCDregularized}).
In the following we present the values of the required $\beta$ functions
and anomalous dimensions, which govern the renormalization constants.

\subsection{$\beta$ functions}
The renormalization of the couplings $\alpha_s, \alphae$, and $\alphafour$ is done
by replacing the bare couplings with the renormalized ones.
As renormalization scheme we choose a modified version of the $\RS$ scheme:
like in Ref.~\cite{Gnendiger:2014nxa} we treat the
multiplicity $\Neps$ of the $\epsilon$-scalars as an initially arbitrary quantity
and subtract divergences of the form $\left(\frac{\Neps}{\epsilon}\right)^n$.
As a consequence, the corresponding $\beta$ functions depend on $\Neps$:
$\overline{\beta}^{i}\equiv\mu^2\frac{\text{d}}{\text{d}\mu^2}\left(\frac{\alpha_i}{4\pi}\right)=
\overline{\beta}^{i}(\alpha_s,\alphae,\alphafour,\Neps)$, 
with $i = s, e, 4\epsilon$.
They are given in Refs.~\cite{Kilgore:2012tb, Gnendiger:2014nxa} and read:
\begin{subequations}
\begin{align}
 \begin{split}
  \overline{\beta}^{s}%(\alpha_s,\alphae,\alphafour)
  = &
    -\Big(\frac{\alphas}{4\pi}\Big)^2 \Bigg[C_A \left(\frac{11}{3}-\frac{\Neps}{6}\right)-\frac{2}{3}N_F\Bigg]\\
   &-\Big(\frac{\alphas}{4\pi}\Big)^3 \Bigg[C_A^2 \left(\frac{34}{3}-\frac{7}{3}\Neps\right)-\frac{10}{3}C_A N_F-2 C_F N_F\Bigg]\\
   &-\Big(\frac{\alphas}{4\pi}\Big)^2 \Big(\frac{\alphae}{4\pi}\Big) \Bigg[C_F N_F\Neps\Bigg]+\mathcal{O}(\alpha^4),
 \end{split}
 \\[15pt]
 \begin{split}
  \overline{\beta}^{e}%(\alpha_s,\alphae,\alphafour)
  = &
  -\Big(\frac{\alphas}{4\pi}\Big) \Big(\frac{\alphae}{4\pi}\Big)\,6\,C_F
  -\Big(\frac{\alphae}{4\pi}\Big)^2 \Bigg[C_A (2-\Neps)+C_F (-4+\Neps)-N_F\Bigg]
  %\\
  %&-\Big(\frac{\alphas}{4\pi}\Big)^2 \Big(\frac{\alphae}{4\pi}\Big) \Bigg[C_A^2 \Big(-\frac{7}{4}+\frac{\Neps}{4}\Big)
  %   +C_A C_F \Big(\frac{61}{3}-\frac{11}{6}\Neps\Big)+C_A N_F\\
  %&\quad\quad\quad\quad\quad\quad\quad+3\,C_F^2-\frac{10}{3}C_F N_F\Bigg]\\
  %&-\Big(\frac{\alphas}{4\pi}\Big) \Big(\frac{\alphae}{4\pi}\Big)^2 \Bigg[-6\,C_A^2+C_A C_F\,(28-11\,\Neps)+C_F^2\,(-32+8 \Neps)-5\,C_F N_F\Bigg]\\
  %&-\Big(\frac{\alphae}{4\pi}\Big)^3 \Bigg[C_A^2 \Big(6-8\,\Neps+\frac{3}{2}\Neps^2\Big)+C_A C_F \Big(-20+24\,\Neps-4\,\Neps^2\Big)+\\
  %&\quad\quad\quad\quad\quad+C_A N_F\Big(-3+\frac{3}{2}\Neps\Big)+C_F^2 \Big(16-16\,\Neps+\frac{9}{4}\Neps^2\Big)+6\,C_F N_F\Bigg]\\
  +\mathcal{O}(\alpha^3).
  \end{split}
\end{align}
\end{subequations}
The renormalization of the quartic coupling
$\left(\alphafour\right)^{\alpha\beta\gamma\delta}_{abcd}$
is more complicated since the tree-level color structure,
$f_{abe}f_{cde}$, is not preserved under renormalization \cite{Jack:1993ws}.
In the case of an SU(3) gauge group one therefore has to introduce three quartic couplings,
$\alpha_{4\epsilon,i}$ with $i=1,2,3$, each of them related to one specific color structure in a
basis of color space. Examples for such a basis are given e.\,g.
in Refs.~\cite{Harlander:2006rj, Harlander:2006xq}.

In the present case of $H\to g g$ the renormalization constant for $\alphafour$ only
appears in diagrams like the third of Fig.~\ref{fig:CTa}. Hence, only the following
contracted $\beta$ function is needed:
\begin{align}
 \begin{split}
 (\overline{\beta}^{4\epsilon})^{\alpha\beta\gamma\delta}_{abcd}\,
 \delta_{ab}^{\phantom{\beta}}\,
 \gtilde_{\alpha\beta}^{\phantom{\beta}}
 =\Bigg\{
 &\Big(\frac{\alphas}{4\pi}\Big)^2 C_A^2 (9+6\,\Neps)\phantom{\bigg]}%\\&
  +\Big(\frac{\alphas}{4\pi}\Big) \Big(\frac{\alphafour}{4\pi}\Big) C_A^2\,(1-\Neps)\,12\phantom{\bigg]}\\&
  +\Big(\frac{\alphae}{4\pi}\Big)^2 \Big[C_A N_F (4-2\,\Neps)+C_F N_F(-8-4\,\Neps)\Big]\phantom{\bigg]}\\&
  +\Big(\frac{\alphae}{4\pi}\Big) \Big(\frac{\alphafour}{4\pi}\Big) C_A N_F\,(1-\Neps)(-4)\phantom{\bigg]}\\&
  +\Big(\frac{\alphafour}{4\pi}\Big)^2 C_A^2\,(1-\Neps)(-7-2\,\Neps)
  \Bigg\}\,\delta_{cd}\,\gtilde^{\gamma\delta}+\mathcal{O}(\alpha^3).\phantom{\bigg]}
 \end{split}
\end{align}
This result is obtained from a direct off-shell calculation.
It agrees with a general result from \cite{Luo:2002ti}.

\subsection{Anomalous dimensions}
For the off-shell calculation of $\Gamma_{HA^\mu A^\nu}$ also
renormalization of the fields and of the gauge parameter $\xi$ is needed.
The renormalization of $\xi$ is fixed by the requirement that the gauge
fixing term does not renormalize: $\xi\to Z_{\Ahat}\xi$.
The anomalous dimensions $\gamma_i=\mu^2\frac{d}{d\mu^2}\text{ln}\,Z_i$
of gluon and $\epsilon$-scalar fields are obtained from a direct off-shell
calculation of the respective two-loop self energies.
Their values up to two-loop level read:
\begin{subequations}
\begin{align}
\notag
\gamma_{\hat{A}}=&
  -\Big(\frac{\alphas}{4\pi}\Big)\Bigg[C_A\left(\frac{13}{6}-\frac{\xi}{2}-\frac{\Neps}{6}\right)-\frac{2}{3}N_F\Bigg]
\\ \notag
 &-\Big(\frac{\alphas}{4\pi}\Big)^2 \Bigg[C_A^2 \left(\frac{59}{8}-\frac{11}{8}\xi-\frac{\xi^2}{4}-\frac{15}{8}\Neps\right)
    -\frac{5}{2} C_A N_F-2\, C_F N_F\Bigg]\\
 &-\Big(\frac{\alphas}{4\pi}\Big) \Big(\frac{\alphae}{4\pi}\Big) C_F N_F \Neps
    +\mathcal{O}(\alpha^3),\phantom{\Bigg]}
 \label{eq:anomalousDimGlu}
\\[15pt] \notag
\gamma_{\Atilde}=&
  -\Big(\frac{\alphas}{4\pi}\Big)C_A(3-\xi)
  -\Big(\frac{\alphae}{4\pi}\Big)\Big[-N_F\Big]
\\ \notag
 &-\Big(\frac{\alphas}{4\pi}\Big)^2 \Bigg[C_A^2 \left(\frac{61}{6}-2\xi-\frac{\xi^2}{4}-\frac{11}{12}\Neps\right)-\frac{5}{3} C_A N_F\Bigg]
  -\Big(\frac{\alphas}{4\pi}\Big) \Big(\frac{\alphae}{4\pi}\Big)\Big[-5\,C_F N_F\Big]
\\ \notag
 &-\Big(\frac{\alphae}{4\pi}\Big)^2 \Bigg[C_A N_F\left(-1+\frac{\Neps}{2}\right)+C_F N_F\left(2+\frac{\Neps}{2}\right)\Bigg]
  -\Big(\frac{\alphafour}{4\pi}\Big)^2 C_A^2\,(1-\Neps)\,\frac{3}{4}\\
 \phantom{\Bigg]}&+\mathcal{O}(\alpha^3).
 \label{eq:anomalousDimEps}
 \end{align}
 \end{subequations}
Setting $\Neps$ and $\alphae$ to zero in Eq.~(\ref{eq:anomalousDimGlu}) yields the well-known
gluon anomalous dimension in \CDR, see e.\,g.~\cite{Larin:1993tp}.
The value of $\gamma_{\tilde{A}}$ agrees with the general result for the 
anomalous dimension of a scalar field \cite{Luo:2002ti},
confirming the point of view that  $\epsilon$-scalars behave like ordinary
scalar fields with multiplicity $\Neps$.

\section{Operator renormalization and mixing in FDH and DRED}
\label{sec:operatorrenormalization}

The second type of counterterm contributions, denoted by
$\Gamma_{H A^\mu A^\nu}^{\text{1LCT,b}}$ and
$\Gamma_{H A^\mu A^\nu}^{\text{2lCT,b}}$,
originates from the necessary renormalization
of the effective Lagrangian (\ref{Leff}), equivalently of the operators
$O_1$ and $\Otilde_1$.
One major difficulty is that multiplicative renormalization of the parameters
$\lambda$ and $\lambda_\epsilon$ is not sufficient since the operators
mix with further operators.
We will show that the full operator mixing involving gauge non-invariant
operators has to be taken into account. The renormalization constants
cannot be predicted from known QCD renormalization constants but need to
be determined from an off-shell calculation.
The general theory of operator mixing in gauge theories and the
classification of gauge invariant and gauge
non-invariant operators has been developed long
ago~\cite{KlubergStern:1974rs,Joglekar:1975nu,Deans:1978wn}.
% Kluberg: Spiridonov's idea, but for integrated operators
% Joglekar: unintegrated operators, BRS/ST invariance and mixing
% Deans: the same, but also S-matrix elements=> always BRS/ST and
% ghost equation used.

In the following we briefly describe operator mixing in the much simpler case of
\CDR\ and then explain the cases of \FDH\  and \DRED, which involve further operators.

\subsection{Operators in CDR}

In \CDR, a useful basis of scalar dimension-4 operators, which is closed under
renormalization, is given in Ref.~\cite{Spiridonov:1984br}:
\begin{subequations}
\label{eq:O1to5}
\begin{align}
O_1&= -\frac{1}{4} \hat{F}^{\mu\nu}_a \hat{F}_{\mu\nu, a}^{\phantom{\mu}},
\phantom{\frac{1}{1}}
%\label{O1}
\\*
O_2&= 0,
\phantom{\frac{1}{1}}\\*
O_3&=  \frac{i}{2}\,\overline{\psi}\,\overleftrightarrow{\slashed{D}}\,\psi,
\phantom{\frac{1}{1}}\\*
O_4&= \Ahat^{\nu}_a(\Dhat^{\mu}\Fhat_{\mu\nu})_a
      -\gs\overline{\psi}\Ahatslashed\psi
      -(\partial^{\mu}\overline{c}_a)(\partial_{\mu}c_{a}),
\phantom{\frac{1}{1}}\\*
O_5&= (D^\mu \partial_\mu \overline{c})_a c_{a}.
\phantom{\frac{1}{1}}
%\label{O5}
\end{align}
\end{subequations}
Operator $O_{1}$ is gauge invariant and related to coupling renormalization;
$O_2=m\overline{\psi}\psi$ in Ref.~\cite{Spiridonov:1984br} and corresponds
to the fermion mass renormalization; we set $m=0$.
All other operators are constrained by BRS invariance and
Slavnov-Taylor identities \cite{KlubergStern:1974rs,Joglekar:1975nu};
operators $O_4$ and $O_5$ are not gauge invariant. 
The basis is chosen such that $O_{3}$, $O_{4}$ and $O_{5}$
are related to field renormalization of
$\psi$, $\hat{A}^\mu$ and $c$, respectively.
In particular, the first two terms of
$O_4$ are generated by applying the functional derivative
\begin{align}
\Ahat_a^\nu(x)\frac{\delta}{\delta\Ahat_a^\nu(x)}
%- \overline{c}_a(x)\frac{\delta}{\delta\overline{c}_a(x)}
\end{align}
on the gauge invariant part of the QCD action; the remaining term is
then required by BRS invariance and the non-renormalization of the
gauge fixing term.\footnote{%
See Refs.~\cite{Joglekar:1975nu,Deans:1978wn} for more details; the
full operator $O_4$ can be obtained from evaluating $W
Y_{\Ahat^\nu_a}\Ahat^\nu_a+W (\partial^\nu\overline{c}_a)A_{\nu,a}$,
where $W$ is the linearized Slavnov-Taylor operator and
$Y_{\Ahat^\nu_a}$ is the source of the BRS transformation of $\Ahat^\nu_a$ in
the functional integral. Since $W$ is nilpotent, this definition shows
that $O_4$ is compatible with BRS invariance and the Slavnov-Taylor
identity and can appear in the operator mixing.}

The operators renormalize as
\begin{align}
O_i &\to Z_{ij}O_{j,\text{bare}},
\end{align}
where $O_{j,\text{bare}}$ arises from $O_j$ by replacing all
parameters and fields by the respective bare quantities. 
Following an elegant proof in Ref.~\cite{Spiridonov:1984br} the
nontrivial  \CDR\ renormalization matrix $Z_{ij}$ can be written in the form
\begin{align}
Z_{ij} &= \delta_{ij}+\mathbbmsl{D}_i\,\text{ln}\mathbbmsl{Z}_{j}.
\label{SpiridonovResult1}
\end{align}
Here, $\mathbbmsl{D}_i$ are derivatives with respect to parameters and $\mathbbmsl{Z}_j$
are combinations of ordinary QCD renormalization constants. As a
result, in particular the renormalization of $Z_{11}$ is given by
\begin{align}
Z_{11} &= 1+\alpha_s\frac{\partial}{\partial \alpha_s}\ln Z_{\alpha_s},
\label{SpiridonovResult2}
\end{align}
with the multiplicative renormalization constant of $\alpha_s$,
$Z_{\alpha_s}$. In this way the renormalization of the parameter
$\lambda$ in the \CDR\ version of $\LeffOp$ is related to the
renormalization of~$\alpha_s$.

\subsection{Operators in FDH and DRED}

In \FDH\ and \DRED, the basis of operators needs to contain additional
terms involving $\epsilon$-scalars. We use a basis constructed
analogously to Eqs.~\eqref{eq:O1to5} from gauge invariant operators and
operators corresponding to field
renormalization. Then there are two kinds of changes: there are
modifications of the operators $O_{3}$ and $O_4$, and there are
additional basis elements. The new basis operators correspond
to the $\epsilon$-scalar kinetic term, $\Otilde_1$, to the new
parameters $\alphae$ and $\alphaFourEps$, $\Otilde_3$ and
$\Otilde_{4\epsilon,i}$, 
and to the field renormalization of $\Atilde^\mu$, $\Otilde_4$. The
notation is chosen such that in all cases $O_j$ and $\Otilde_j$ have
a similar structure:
\begin{subequations}
\label{eq:O1toO5DRED}
\begin{align}
O_1&= -\frac{1}{4} \hat{F}^{\mu\nu}_a \hat{F}_{\mu\nu, a}^{\phantom{\mu}},
\phantom{\frac{1}{\hat{A}}}
%\label{O1DRED}
\\
O_2&= 0,
\phantom{\frac{1}{\hat{A}}}
\\
O_3&=  \frac{i}{2}\,\overline{\psi}\,\overleftrightarrow{\slashed{D}}\,\psi
  -\ge \overline{\psi} {\Atildeslashed} \psi,
\phantom{\frac{1}{\hat{A}}}
\\
O_4&= \Ahat^{\nu}_a(\Dhat^{\mu}\Fhat_{\mu\nu})_a
  +\gs f_{abc}(\partial^\mu\Atilde^{\nu}_{a})\Ahat_{\mu, b}\Atilde_{\nu, c}
  -\gs\overline{\psi}{\Ahatslashed}\psi-\left(\partial^\mu \overline{c}_a\right)\left(\partial_\mu c_a\right),
\phantom{\frac{1}{\hat{A}}}
\\
O_5&= (\Dhat^\mu \partial_\mu \overline{c}_a)c_a,
\phantom{\frac{1}{\hat{A}}}
\\[15pt]
\Otilde_{1}&= -\frac{1}{2}(\hat{D}^\mu\Atilde^\nu)_a(\hat{D}_\mu\Atilde_\nu)_a,
\phantom{\frac{1}{\hat{A}}}
\\
\Otilde_{3}&= \ge \overline{\psi} \Atildeslashed \psi,
\phantom{\frac{1}{\hat{A}}}
\\
\Otilde_{4}&= \Atilde^\nu_a (\Dhat^\mu \Dhat_\mu\Atilde_\nu)_a,
\phantom{\frac{1}{\hat{A}}}
\\
\Otilde_{4\epsilon,i} & = {\cal O}(\Atilde^4).
\phantom{\frac{1}{\hat{A}}}
%\label{O4tDRED}
\end{align}
\end{subequations}
Since we consider massless QCD there is no $\epsilon$-scalar mass term.
Like in Eq.\ (\ref{Leff}), operators involving four
$\epsilon$-scalars are not needed explicitly. 

This set of operators differs in a crucial way from the
\CDR\ case. The difference between operators $\Otilde_1$ and $\Otilde_4$
is related to the total derivative $\Box\Atilde^\mu\Atilde_\mu$.
Hence, the basis for space-time
integrated operators (zero-momentum insertions) does not coincide with
the one for non-integrated operators (non-vanishing momentum
insertions). 
As discussed by Spiridonov in Ref.\ \cite{Spiridonov:1984br}, in such a case
his method cannot be used. Therefore, in \FDH\ and \DRED\ it is
not possible to derive 
complete results for the operator mixing  analogous to
Eqs.~(\ref{SpiridonovResult1}) and (\ref{SpiridonovResult2}). 

This implies two difficulties: First, the two-loop renormalization of
$\Otilde_1$ and the corresponding parameter $\lambda_\epsilon$ cannot
be obtained from a priori
known two-loop QCD renormalization constants but need to be determined
from an explicit two-loop off-shell calculation. Second, the off-shell
Green functions get contributions from unphysical, gauge non-invariant
operators, so the full operator mixing needs to be taken into account.

We have carried out the explicit one-loop calculations to obtain all
required one-loop results for $Z_{1j}$ and $Z_{\tilde{1}j}$.
The results are
\begin{subequations}
\label{eq:dZij}
\begin{align}
\delta Z_{11}^{\text{1L}} &=\left(\frac{\alphas}{4\pi} \right)
  \left[\Big(-\frac{11}{3}+\frac{\Neps}{6}\,\Big)C_A+\frac{2}{3}N_F\right]\frac{1}{\epsilon},
\label{Z11}
\\*
\delta Z_{\tilde{1}1}^{\text{1L}} &= 0,\phantom{\frac{\alphas}{4\pi}}
\label{Zt11}
\\
\delta Z_{1\tilde{1}}^{\text{1L}} &= 0,\phantom{\frac{\alphas}{4\pi}}
\label{Z11t}
\\*
\delta Z_{\tilde{1}\tilde{1}}^{\text{1L}} &=\bigg[
  \Big(\frac{\alphas}{4\pi}\Big)(-3)C_A
  +\Big(\frac{\alphae}{4\pi}\Big) N_F
  -\Big(\frac{\alphafour}{4\pi} \Big)(1-\Neps)C_A
  \bigg]\frac{1}{\epsilon},
  \label{Z1t1t}
\\
\delta Z_{13}^{\text{1L}} &= 0,\phantom{\frac{\alphas}{4\pi}}
\\*
\delta Z_{\tilde{1}3}^{\text{1L}} &= \left(\frac{\alphae}{4\pi} \right)
  \frac{\Neps}{2}C_F\frac{1}{\epsilon},
\\
%Z_{1\tilde{3}}^{\text{1L}} &= \dots \text{unknown and uninteresting},
%\phantom{\frac{\alphas}{4\pi}}
%\\*
%Z_{\tilde{1}\tilde{3}}^{\text{1L}} &= \dots,\phantom{\frac{\alphas}{4\pi}}
%\\
\delta Z_{14}^{\text{1L}} &= \left(\frac{\alphas}{4\pi} \right)
  \frac{3}{4}C_A\frac{1}{\epsilon},
\\*
\delta Z_{\tilde{1}4}^{\text{1L}} &= 0,\phantom{\frac{\alphas}{4\pi}}
\\
\delta Z_{1\tilde{4}}^{\text{1L}} &= \left(\frac{\alphas}{4\pi} \right)
  \Big(-\frac{3}{2}\Big)C_A\frac{1}{\epsilon},
\\*
\delta Z_{\tilde{1}\tilde{4}}^{\text{1L}} &= \left(\frac{\alphas}{4\pi} \right)
  \frac{1}{2}(3-\xi)C_A\frac{1}{\epsilon},
\\
\delta Z_{15}^{\text{1L}} &= 0,\phantom{\frac{\alphas}{4\pi}}
\\*
\delta Z_{\tilde{1}5}^{\text{1L}} &= 0.\phantom{\frac{\alphas}{4\pi}}
%\label{Zt15}
\end{align}
\end{subequations}
Renormalization constants involving operators $\Otilde_3$ or
$\Otilde_{4\epsilon,i}$ are not needed for the calculations in the
present paper.
The renormalization constants (\ref{Z11})-(\ref{Z1t1t}) agree with those
given in Ref.~\cite{Gnendiger:2014nxa}. The only gauge-dependent quantity is
$Z_{\tilde{1}\tilde{4}}^{\text{1L}}$. This is due to the fact that operator
$\Otilde_4$ is related to the field renormalization of the $\epsilon$-scalars.
In all other renormalization constants related to field renormalization the
gauge-dependent parts incidentally cancel out.

With these results the bare effective Lagrangian can be written as
\begin{align}
\LeffOp^{\text{bare}} &=
H \sum_{j}\left(
\lambda\, Z_{1j}O_{j,\text{bare}}
+
\lambda_\epsilon\, Z_{\tilde{1}j}O_{j,\text{bare}}^{\phantom{I}}
\right),
\end{align}
where the sum runs over all operators in Eqs.~\eqref{eq:O1toO5DRED}.
Sometimes it is useful to write this using multiplicative renormalization
constants for $\lambda$ and $\lambda_\epsilon$ as
\begin{align}
\LeffOp^{\text{bare}} &=
Z_{\lambda} \lambda H O_{1,\text{bare}}
+
Z_{\lambda_\epsilon} \lambda_\epsilon H O_{\tilde{1},\text{bare}}
+\ldots,
\label{Zlambda}
\end{align}
suppressing operators not present at tree level, such that 
$\lambda Z_{\lambda}  =
\lambda Z_{11} +
\lambda_\epsilon  Z_{\tilde{1}1} $ and similar for
$Z_{\lambda_\epsilon}$.

\begin{figure}[t]
\begin{center}
\scalebox{.70}{
\begin{picture}(135,90)(0,0)
\Vertex(25,45){2}
\Text(25,37.5)[b]{\scalebox{2}{\ding{53}}}
\Vertex(62.5,62.5){2}
\Vertex(62.5,27.5){2}
\DashLine(0,45)(25,45){2}
\ArrowLine(62.5,62.5)(25,45)
\ArrowLine(25,45)(62.5,27.5)
\ArrowLine(62.5,27.5)(62.5,62.5)
\DashLine(62.5,62.5)(100,80){4}
\DashLine(62.5,27.5)(100,10){4}
\Text(5,50)[b]{\scalebox{1.42}{$O_3$}}
\end{picture}
\quad
\begin{picture}(135,90)(0,0)
\Vertex(30,45){2}
%\Line(22,53)(38,37)
%\Line(22,37)(38,53)
\Text(30,37.5)[b]{\scalebox{2}{\ding{53}}}
\Vertex(60,65){2}
\ArrowArc(47.5,52)(18,40,200)
\ArrowArc(43,59)(18,240,15)
%\GlueArc(52,45)(20,65,185){3}{4}
%\DashCArc(39,65)(22,250,5){4}
\DashLine(0,45)(30,45){2}
\DashLine(30,45)(100,10){4}
\DashLine(60,65)(100,90){4}
\Text(10,50)[b]{\scalebox{1.42}{$O_3$}}
\end{picture}
}
\quad
\scalebox{.70}{
\begin{picture}(135,90)(0,0)
\Vertex(25,45){2}
%\Line(17,53)(33,37)
%\Line(17,37)(33,53)
\Text(25,37.5)[b]{\scalebox{2}{\ding{53}}}
\Vertex(62.5,62.5){2}
\Vertex(62.5,27.5){2}
\DashLine(0,45)(25,45){2}
\Gluon(25,45)(62.5,62.5){4}{4}
\Gluon(25,45)(62.5,27.5){4}{4}
\DashLine(62.5,62.5)(100,80){4}
\DashLine(62.5,27.5)(100,10){4}
\DashLine(62.5,27.5)(62.5,62.5){4}
\Text(5,50)[b]{\scalebox{1.42}{$O_4$}}
\end{picture}
\quad
\begin{picture}(135,90)(0,0)
\Vertex(30,45){2}
%\Line(22,53)(38,37)
%\Line(22,37)(38,53)
\Text(30,37.5)[b]{\scalebox{2}{\ding{53}}}
\Vertex(60,65){2}
\GlueArc(47.5,52)(18,40,200){4}{4}
\DashCArc(43,59)(18.5,240,15){4}
\DashLine(0,45)(30,45){2}
\DashLine(30,45)(100,10){4}
\DashLine(60,65)(100,90){4}
\Text(10,50)[b]{\scalebox{1.42}{$O_4$}}
\end{picture}
}
\\[15pt]
\scalebox{.70}{
\begin{picture}(135,90)(0,0)
\Vertex(25,45){2}
%\Line(17,53)(33,37)
%\Line(17,37)(33,53)
\Text(25,37.5)[b]{\scalebox{2}{\ding{53}}}
\Vertex(62.5,62.5){2}
\Vertex(62.5,27.5){2}
\DashLine(0,45)(25,45){2}
\DashLine(62.5,62.5)(25,45){4}
\DashLine(25,45)(62.5,27.5){4}
\Gluon(62.5,27.5)(62.5,62.5){4}{4}
\DashLine(62.5,62.5)(100,80){4}
\DashLine(62.5,27.5)(100,10){4}
\Text(5,50)[b]{\scalebox{1.42}{$\Otilde_4$}}
\end{picture}
\quad
\begin{picture}(135,90)(0,0)
\Vertex(30,45){2}
%\Line(22,53)(38,37)
%\Line(22,37)(38,53)
\Text(30,37.5)[b]{\scalebox{2}{\ding{53}}}
\Vertex(60,65){2}
\GlueArc(47.5,52)(18,40,200){4}{4}
\DashCArc(43,59)(18.5,240,15){4}
%\GlueArc(52,45)(20,65,185){3}{4}
%\DashCArc(39,65)(22,250,5){4}
\DashLine(0,45)(30,45){2}
\DashLine(30,45)(100,10){4}
\DashLine(60,65)(100,90){4}
\Text(10,50)[b]{\scalebox{1.42}{$\Otilde_4$}}
\end{picture}
}
\quad
\scalebox{.70}{
\begin{picture}(135,90)(0,0)
\Vertex(25,45){2}
%\Line(17,53)(33,37)
%\Line(17,37)(33,53)
\Text(25,37.5)[b]{\scalebox{2}{\ding{53}}}
\Vertex(62.5,62.5){2}
\Vertex(62.5,27.5){2}
\DashLine(0,45)(25,45){2}
\DashArrowLine(62.5,62.5)(25,45){2}
\DashArrowLine(25,45)(62.5,27.5){2}
\DashArrowLine(62.5,27.5)(62.5,62.5){2}
\Gluon(62.5,62.5)(100,80){4}{4}
\Gluon(62.5,27.5)(100,10){4}{4}
\Text(5,50)[b]{\scalebox{1.42}{$O_5$}}
\end{picture}
\quad
\begin{picture}(135,90)(0,0)
\Vertex(30,45){2}
%\Line(22,53)(38,37)
%\Line(22,37)(38,53)
\Text(30,37.5)[b]{\scalebox{2}{\ding{53}}}
\Vertex(60,65){2}
\DashArrowArc(47.5,52)(18,40,200){2}
\DashArrowArc(43,59)(18.5,240,15){2}
\DashLine(0,45)(30,45){2}
\Gluon(60,65)(100,90){4}{4}
\Gluon(100,10)(30,45){4}{8}
\Text(10,50)[b]{\scalebox{1.42}{$O_5$}}
\end{picture}
}
\caption{\label{fig:operatorRenSample}
Sample one-loop counterterm diagrams originating from operators
$O_3$, $O_4$, $\Otilde_4$ and $O_5$.
}
\end{center}
\end{figure}
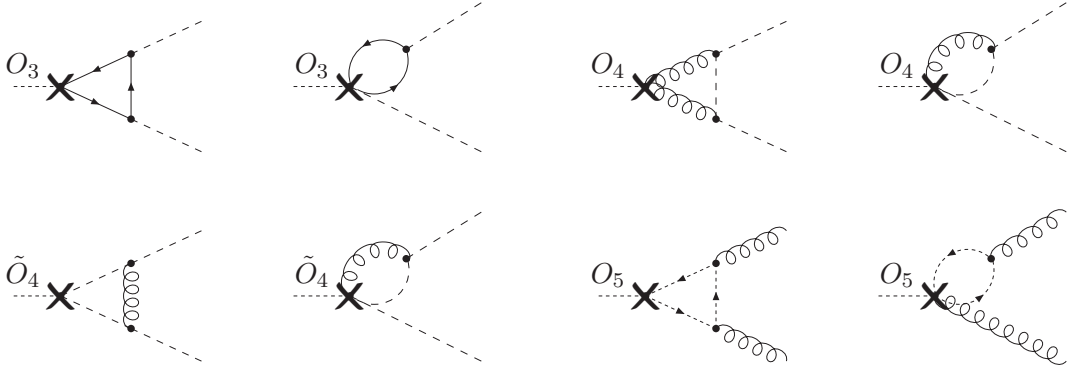

The one-loop counterterm effective Lagrangian involving the
renormalization constants of Eqs.~\eqref{eq:dZij} is
then given by
\begin{align}
\LeffOp^{\text{1LCT}} &=
H\sum_{j}\left(
\lambda\,\delta Z_{1j}^{\text{1L}}O_{j}^{\phantom{\text{1L}}}
+
\lambda_{\epsilon}\,\delta Z_{\tilde{1}j}^{\text{1L}}
O_{j\phantom{\tilde{1}}}^{\phantom{\text{1L}}}
\right).
\end{align}
We have now all ingredients for the one-loop counterterm diagrams $\Gamma_{H
  A^\mu A^\nu}^{\text{1LCT,b}}$ 
relevant for the computation of
$H\to gg$, where the gluons are either $D$-dimensional gauge fields or
$\epsilon$-scalars. These counterterm contributions arise from
one-loop counterterm diagrams with one insertion of 
$\LeffOp^{\text{1LCT}}$. Sample diagrams are given in
Fig.\ \ref{fig:operatorRenSample}. They show insertions of
operators $O_3$, $O_4$, $\Otilde_4$ and $O_5$.
Feynman rules for operator insertions are given in
appendix~\ref{sec:appendix_B}.

The calculation shows that all these operators 
generate non-vanishing contributions to
$\Gamma_{H A^\mu  A^\nu}^{\text{1LCT,b}}$.
However, in the extraction of the form factors and two-loop
renormalization constants to be discussed in the next section
there are cancellations, and $O_4$ is the only new operator which
contributes.

\section{Two-loop renormalization constants of $\lambda$ and $\lambda_\epsilon$}
\label{sec:CT2Lb}

%At the two-loop level, the structure of $\LeffOp^{\text{1LCT}}$ is
%unchanged, but for our purposes we only need the renormalization
%constants  $\delta Z^{\text{2L}}_{\lambda}$ and $\delta
%Z^{\text{2L}}_{\leps}$, corresponding to a two-loop multiplicative
%renormalization of the tree-level Lagrangian (\ref{Leff}). Using the
%general approach of the previous section, we have the identification
%\begin{align}
%\delta Z^{\text{2L}}_{\lambda} \lambda &=
%Z_{11}^{\text{2L}}\lambda +
%  Z_{\tilde{1}1}^{\text{1L}}\lambda_\epsilon
%&
%\delta Z^{\text{2L}}_{\lambda} \lambda &=
%Z_{1\tilde{1}}^{\text{2L}}\lambda +
%  Z_{\tilde{1}\tilde{1}}^{\text{1L}}\lambda_\epsilon
%.
%\end{align}
%\marginpar{necessary? correct?}
Putting together the results from the previous three sections it is possible to calculate the
two-loop renormalization constants $\delta Z^{\text{2L}}_{\lambda}$
and $\delta Z^{\text{2L}}_{\leps}$ appearing in Eq.~(\ref{Zlambda}).
They can be obtained from a complete  off-shell two-loop calculation and the requirement
that the corresponding Green-functions are UV finite after renormalization:
\begin{align}
  \Gamma_{H A^\mu A^\nu}^{\text{2L}}
 +\Gamma_{H A^\mu A^\nu}^{\text{1LCT,a}}
 +\Gamma_{H A^\mu A^\nu}^{\text{2LCT,a}}
 +\Gamma_{H A^\mu A^\nu}^{\text{1LCT,b}}
 +\Gamma_{H A^\mu A^\nu}^{\text{2LCT,b}}\bigg|_{\text{UV div.}}^{\text{off-shell}}= 0.
 \label{deltaLambda}
\end{align}
All ingredients except the last term are computed in the
previous sections, and Eq.~\eqref{deltaLambda} is then used to extract $\delta
Z^{\text{2L}}_{\lambda}$ and $\delta Z^{\text{2L}}_{\leps}$.
The result for $\delta Z^{\text{2L}}_{\lambda}$ is:
\begin{align}
\begin{split}
 \delta Z^{\text{2L}}_{\lambda} = &
 \Big(\frac{\alphas}{4\pi}\Big)^2 \Bigg\{
 C_A^2 \Bigg[
    \frac{\frac{121}{9}-\frac{11}{9}\Neps+\frac{\Neps^2}{36}}{\epsilon^2}
   +\frac{-\frac{34}{3}+\frac{7}{3}\Neps}{\epsilon}
   \Bigg]
 \\&\quad\quad\quad\quad
 +C_A N_F \Bigg[
    \frac{-\frac{44}{9}+\frac{2}{9}\Neps}{\epsilon^2}
   +\frac{10}{3\epsilon }
 \Bigg]
    +C_F N_F\frac{2}{\epsilon }+N_F^2\frac{4}{9 \epsilon ^2}\Bigg\}\\
 &+\Big(\frac{\alphas}{4\pi}\Big) \Big(\frac{\alphae}{4\pi}\Big)
 C_F N_F\frac{\left(-1-\frac{\leps}{\lambda }\right)\Neps}{2 \epsilon }.
 \label{eq:dZlambda}
\end{split}
\end{align}
Since the
off-shell calculations have been done numerically with the help of
FIESTA~\cite{Smirnov:2008py} the analytical expressions have been
obtained by rounding to a least
common denominator. The numerical uncertainty is less than
$\frac{1}{72}$ for the terms of the order
$\mathcal{O}(\epsilon^{-2})$ and $\frac{1}{6}$ for the terms of the
order $\mathcal{O}(\epsilon^{-1})$. 

Result (\ref{eq:dZlambda}) is not new; it agrees with Ref.~\cite{Gnendiger:2014nxa},
where it has been obtained using Spiridonov's method. The recalculation serves as a test of the
setup and the results given in the previous sections. At the same time
a comparison with Ref.~\cite{Gnendiger:2014nxa}  confirms that
Eq.~(\ref{eq:dZlambda}) is actually exactly correct, in spite of
numerical uncertainties. 

In the same way, we obtain the renormalization constant $\delta Z^{\text{2L}}_{\leps}$:
\begin{align}
\notag
 \delta Z^{\text{2L}}_{\leps} = 
 & \Big(\frac{\alphas}{4\pi}\Big)^2\Bigg\{
 C_A^2 \Bigg[
    \frac{\frac{49}{4}+\frac{5}{4}\Neps}{\epsilon ^2}
   +\frac{-\frac{113}{24}+\frac{71}{24}\Neps+\frac{\lambda}{\leps}\Big(2-\frac{\Neps}{2}\Big)}{\epsilon }
   \Bigg]
 +C_A N_F \Bigg[
   -\frac{1}{\epsilon^2}
   +\frac{\frac{5}{6}-2\frac{\lambda }{\leps}}{\epsilon}
   \Bigg]
 \Bigg\}
\\ \notag
 &+\Big(\frac{\alphas}{4\pi}\Big)  \Big(\frac{\alphae}{4\pi}\Big) \Bigg\{
 C_A N_F \Bigg[
   -\frac{3}{\epsilon ^2}
   +\frac{\frac{3}{2}+3\frac{\lambda }{\leps}}{\epsilon }
   \Bigg]
 +C_F N_F \Bigg[
    -\frac{3}{\epsilon ^2}
    +\frac{\frac{5}{2}-3\frac{\lambda }{\leps}}{\epsilon }
    \Bigg]\Bigg\}
\\ \notag
 &+\Big(\frac{\alphas}{4\pi}\Big)  \Big(\frac{\alphafour}{4\pi}\Big)
 C_A^2\,(1-\Neps)\Bigg[
    \frac{6}{\epsilon ^2}
   +\frac{-4-3\frac{\lambda}{\leps}}{\epsilon }
   \Bigg]
\\ \notag
 &+\Big(\frac{\alphae}{4\pi}\Big)^2\Bigg\{
 C_A N_F\frac{-\frac{3}{2}+\frac{3}{4}\Neps}{\epsilon }
+C_F N_F \Bigg[
   -\frac{3\Neps}{2\,\epsilon ^2}
   +\frac{3-\frac{7}{4}\Neps}{\epsilon }
   \Bigg]
+\frac{N_F^2}{
   \epsilon ^2}
   \Bigg\}
\\ \notag
 &+\Big(\frac{\alphae}{4\pi}\Big) \Big(\frac{\alphafour}{4\pi}\Big)
 C_A N_F\,(1-\Neps)\Bigg[
   -\frac{2}{\epsilon^2}
   +\frac{3}{2\epsilon}
   \Bigg]
\\
 &+\Big(\frac{\alphafour}{4\pi}\Big)^2 C_A^2\,(1-\Neps) \Bigg[
    \frac{-\frac{5}{4}-\Neps}{\epsilon^2}
   +\frac{15}{8\epsilon}
   \Bigg].
 \label{eq:dZlambdaeps}
\end{align}
Compared to Eq.~(\ref{eq:dZlambda}) this result is more complicated and includes all combinations
of the three couplings $\alphas, \alphae$ and $\alphafour$.
This result is new; as described in
Sec.\ \ref{sec:operatorrenormalization} it cannot be obtained using
Spiridonov's method. The numerical uncertainty is less than
$\frac{1}{48}$ for all terms. A forthcoming comparison with a prediction of the
infrared structure of $H\to \gtilde\gtilde$ will confirm that
expression (\ref{eq:dZlambdaeps}) is exactly correct \cite{IRstructure}. 

\section{UV renormalized form factors of gluons and $\epsilon$-scalars}
\label{sec:onshell_results}
Now that all renormalization constants are known it is possible
to calculate the two-loop form factors of gluons and $\epsilon$-scalars
in the \FDH\ and \DRED\ scheme.
We present the results in two ways: First, we give results with
independent couplings needed to determine the IR anomalous dimensions
of gluons and $\epsilon$-scalars; second, we give simplified
results, where all couplings are set equal. These can be viewed as the
final results for the UV renormalized but IR regularized form factors.
We give them including higher orders in the $\epsilon$-expansion.

\subsection{Results for independent couplings}
\label{sec:onshell_results_complete}
The UV renormalized but IR divergent form factor for $H\to \ghat\ghat$
in \DRED\ is given at the one-loop and two-loop level by
\begin{align}
\notag
\bar{F}^{\text{1L}}_{\ghat} & (\alphas,\HiggsEps/\HiggsGlu,\Neps)
\\ \notag
=&\Big(\frac{\alphas}{4\pi}\Big)
\Bigg\{C_A \Bigg[
  -\frac{2}{\epsilon ^2}
  +\frac{-\frac{11}{3}+\frac{\Neps}{6}}{\epsilon }
  +\frac{\pi ^2}{6}+\frac{\leps}{\lambda }\Neps
  +\epsilon  \Big(-2+\frac{14}{3}\zeta (3)+3\frac{\leps}{\lambda }\Neps\Big)%\\
  %&\quad\quad\quad\quad\quad
  %+\epsilon ^2 \Big(-6+\frac{47}{720}\pi ^4
  %+\frac{\leps}{\lambda}\Neps\Big(7-\frac{\pi^2}{12}\Big)\Big)
  \Bigg]
+\frac{2 N_F}{3 \epsilon }\Bigg\}
\\&+\mathcal{O}(\epsilon^2),\phantom{\Bigg\}}
\label{eq:gluonFF1}
\\[15pt]
%\begin{split}
\notag
\bar{F}^{\text{2L}}_{\ghat} & (\alphas,\alphae,\HiggsEps/\HiggsGlu,\Neps)
\\ \notag
  =&\Big(\frac{\alphas}{4\pi}\Big)^2 \Bigg\{C_A^2 \Bigg[
       \frac{2}{\epsilon ^4}
      +\frac{\frac{77}{6}-\frac{7}{12}\Neps}{\epsilon^3}
      +\frac{\frac{175}{18}-\frac{\pi ^2}{6}-\Neps\Big(1+2\frac{\leps}{\lambda}\Big)+\frac{\Neps^2}{36}}{\epsilon ^2}
\\ \notag
      &\quad\quad\quad\quad\quad\quad
      +\frac{-\frac{238}{27}-\frac{11}{36}\pi ^2-\frac{25}{3}\zeta (3)
      +\Neps\Big(\frac{49}{27}+\frac{\pi^2}{72}-\frac{29}{3}\frac{\leps}{\lambda }\Big)
      +\frac{1}{6}\frac{\leps}{\lambda }\Neps^2}{\epsilon}%\\
      %&\quad\quad\quad\quad\quad\quad
      %+\frac{8237}{162}+\frac{67}{36}\pi^2-33\zeta(3)-\frac{7}{60}\pi^4
      %+\Neps\Big(-\frac{1003}{324}-\frac{\pi^2}{9}+\frac{3}{2}\zeta(3)\Big)\\
      %&\quad\quad\quad\quad\quad\quad
      %+\frac{\leps}{\lambda}\Neps\Big(-5+\frac{\pi^2}{3}+\frac{1}{2}\Neps\Big)
      \Bigg]
\\ \notag
      &\quad\quad\quad\quad
   +C_A N_F \Bigg[
      -\frac{7}{3 \epsilon ^3}
      +\frac{-\frac{13}{3}+\frac{2}{9}\Neps}{\epsilon ^2}
      +\frac{\frac{64}{27}+\frac{\pi ^2}{18}+\frac{2}{3}\frac{\leps}{\lambda }\Neps}{\epsilon}%\\
      %&\quad\quad\quad\quad\quad\quad\quad\quad\,\
      %-\frac{1024}{81}-\frac{5}{18}\pi^2-2\zeta(3)+2\frac{\leps}{\lambda}\Neps
      \Bigg]
    %\\&\quad\quad\quad\quad
   +C_F N_F%\Bigg[
       \frac{1}{\epsilon}%-\frac{67}{6}+8\zeta(3)
       %\Bigg]
   +\frac{4 N_F^2}{9 \epsilon ^2}\Bigg\}
\\
&-\Big(\frac{\alphas}{4\pi}\Big)\Big(\frac{\alphae}{4\pi}\Big)%\Bigg\{
    %C_A N_F\frac{17}{4}\frac{\leps}{\lambda}
    C_F N_F%\Bigg[
       \frac{\Neps}{2\epsilon}%+\frac{1}{4}\frac{\leps}{\lambda}
       %\Bigg]
    %\Bigg\}\\
%&+\Big(\frac{\alphas}{4\pi}\Big)\Big(\frac{\alphafour}{4\pi}\Big)
 %  C_A^2\frac{\leps}{\lambda}\Neps 2 (1-\Neps)
    +\mathcal{O}(\epsilon^0).\phantom{\Bigg\}}
    \label{eq:gluonFF2}
 %\end{split}
 \end{align}
As mentioned in the beginning the $\ghat$ form factor in \DRED\ is
identical to the gluon form factor in \FDH, and Eq.~(\ref{eq:gluonFF2})
agrees with the result given in Ref.~\cite{Gnendiger:2014nxa}.

Since there are no external $\epsilon$-scalars in diagrams related to the gluon
form factor internal $\epsilon$-scalars have to be part of a closed $\epsilon$-scalar loop
or have to couple to a closed fermion loop.
Hence, the effective coupling $\leps$ always appears together with at
least one power of $\Neps$ in Eqs.~(\ref{eq:gluonFF1}) and (\ref{eq:gluonFF2}).
 
The $\epsilon$-scalar form factor for $H\to\gtilde\gtilde$ in
\DRED\ is given by
 \begin{align}
 \notag
   \bar{F}^{\text{1L}}_{\gtilde} & (\alphas,\alphae,\alphafour,\HiggsGlu/\HiggsEps,\Neps)
 \\ \notag
  =&\Big(\frac{\alphas}{4\pi}\Big)
    C_A \Bigg[
      -\frac{2}{\epsilon ^2}
      -\frac{4}{\epsilon }
      -2+\frac{\pi ^2}{6}+2\frac{\lambda }{\leps}
      +\epsilon  \Big(-4+\frac{\pi^2}{12}+\frac{14}{3}\zeta (3)+4\frac{\lambda}{\leps}\Big)  
      \Bigg]\\
  &+\Big(\frac{\alphae}{4\pi}\Big)\frac{N_F}{\epsilon }
   +\Big(\frac{\alphafour}{4\pi}\Big)
    C_A\,(1-\Neps)\Bigg[
    2%-2\Neps
    +\epsilon\Big(4-\frac{\pi ^2}{12}
    \Big)
    \Bigg]%\\&
    +\mathcal{O}(\epsilon^2),\phantom{\Bigg\}}
\\[15pt] \notag
 \bar{F}^{\text{2L}}_{\gtilde} & (\alphas,\alphae,\alphafour,\HiggsGlu/\HiggsEps,\Neps)
 \\ \notag
  =&\Big(\frac{\alphas}{4\pi}\Big)^2 \Bigg\{C_A^2 \Bigg[
           \frac{2}{\epsilon ^4}
          +\frac{\frac{27}{2}-\frac{\Neps}{4}}{\epsilon^3}
          +\frac{\frac{281}{18}-\frac{\pi ^2}{6}-\frac{\Neps}{9}-4\frac{\lambda }{\leps}}{\epsilon^2}
\\ \notag
          &\quad\quad\quad\quad\quad\quad
          +\frac{\frac{469}{216}-\frac{5}{12}\pi ^2-\frac{25}{3}\zeta (3)+\Neps\Big(\frac{233}{216}+\frac{\pi^2}{24}\Big)-16\frac{\lambda }{\leps}}{\epsilon}%\\
          \Bigg]
\\ \notag
  &\quad\quad\quad\quad+C_A N_F \Bigg[
         -\frac{1}{\epsilon ^3}-\frac{7}{9 \epsilon ^2}
         +\frac{\frac{113}{54}+\frac{\pi ^2}{6}}{\epsilon }
         \Bigg]\Bigg\}
\\ \notag
  &+\Big(\frac{\alphas}{4\pi}\Big)\Big(\frac{\alphae}{4\pi}\Big) \Bigg\{C_A N_F \Bigg[
        -\frac{2}{\epsilon ^3}-\frac{4}{\epsilon^2}
        +\frac{-2-\frac{\pi ^2}{6}+2\frac{\lambda }{\leps}}{\epsilon}
        \Bigg]
      +C_F N_F \Bigg[
        -\frac{3}{\epsilon ^2}+\frac{5}{2 \epsilon}
        \Bigg]\Bigg\}
\\ \notag
  &+\Big(\frac{\alphas}{4\pi}\Big)\Big(\frac{\alphafour}{4\pi}\Big) C_A^2\,(1-\Neps)\Bigg[
        -\frac{4}{\epsilon ^2}
        +\frac{-16+\frac{\pi ^2}{6}}{\epsilon}
        \Bigg]
\\ \notag
  &+\Big(\frac{\alphae}{4\pi}\Big)^2 \Bigg\{C_A N_F \Bigg[
         \frac{-1+\frac{\Neps}{2}}{\epsilon ^2}
        +\frac{\frac{1}{2}-\frac{\Neps}{4}}{\epsilon}
        \Bigg]
      +C_F N_F \Bigg[
         \frac{2-\frac{\Neps}{2}}{\epsilon^2}
        +\frac{-1-\frac{\Neps}{4}}{\epsilon}
        \Bigg]
      +\frac{N_F^2}{\epsilon ^2}\Bigg\}
\\ \notag
  &+\Big(\frac{\alphae}{4\pi}\Big)\Big(\frac{\alphafour}{4\pi}\Big)C_A N_F\,(1-\Neps)\,%\Bigg[
       \frac{2}{\epsilon}\phantom{\Bigg\}}
      %+\left(-\frac{73}{4}+\frac{\pi^2}{12}\right)
  %\Bigg]
\\
  &+\Big(\frac{\alphafour}{4\pi}\Big)^2 C_A^2%\Bigg[
  (1-\Neps)\frac{-3}{8\epsilon}
  %-\frac{139}{4}-\frac{11}{24}\pi^2
  %+\Neps\left(\frac{107}{4}+\frac{13}{24}\pi^2\right)
  %+\Neps^2\left(8-\frac{\pi^2}{12}\right)
  %\Bigg]\\&
  +\mathcal{O}(\epsilon^0).\phantom{\Bigg\}}
\end{align}
Compared to Eqs.~(\ref{eq:gluonFF1}) and (\ref{eq:gluonFF2}) the result with external
$\epsilon$-scalars is more complicated and includes all combinations of the couplings
$\alphas, \alphae$ and $\alphafour$.
In this result, like in all previous results, the evanescent coupling $\alphae$
appears always together with at least one power of $N_F$ and the quartic coupling
$\alphafour$ is always accompanied by a factor $(1-\Neps)$.

\subsection{Results for equal couplings}

During the renormalization process the couplings $\alphas$, $\alphae$, $\alphafour$
and $\lambda$, $\leps$ have to be distinguished.
After renormalization they can be set equal, giving a simpler form of
the final result.\footnote{%
If the results of Sec.~\ref{sec:onshell_results_complete} were not desired
for independent couplings, the genuine two-loop diagrams could have been computed
in a simpler way, with all couplings set equal from the beginning --- this is
what is done in many applications of \FDH\ and \DRED\ in the literature.}
The results for $\Neps=2\epsilon$ at the one(two)-loop level up
to order $\mathcal{O}(\epsilon^4)$ ($\mathcal{O}(\epsilon^2)$) then read:
\begin{align}
\notag
\bar{F}^{\text{1L}}_{\ghat} = \ & \Big(\frac{\alphas}{4\pi}\Big)
\Bigg\{
 C_A \Bigg[
   -\frac{2}{\epsilon ^2}
   -\frac{11}{3 \epsilon }
   +\frac{1}{3}
   +\frac{\pi ^2}{6}
   +\epsilon\frac{14}{3}\zeta (3)
   +\epsilon ^2\frac{47}{720}\pi ^4
   \\&\quad\quad\quad\quad\quad\notag
   +\epsilon^3 \Big(
      \frac{62}{5}\zeta (5)
     -\frac{7}{18}\pi ^2 \zeta (3)
     \Big)
   +\epsilon^4 \Big(
      \frac{949}{60480}\pi ^6
     -\frac{49}{9}\zeta(3)^2
     \Big)
   \Bigg]
 \\ & \quad\quad\quad
 +\frac{2 N_F}{3 \epsilon}
 \Bigg\}
 +\mathcal{O}(\epsilon^5),
\\[15pt] \notag
  \bar{F}^{\text{2L}}_{\ghat}
  = \ &\Big(\frac{\alphas}{4\pi}\Big)^2
  \Bigg\{
C_A^2 \Bigg[
     \frac{2}{\epsilon ^4}
    +\frac{77}{6 \epsilon ^3}
    +\frac{
       \frac{77}{9}
      -\frac{\pi ^2}{6}
      }{\epsilon ^2}
    +\frac{
      -\frac{400}{27}
      -\frac{11}{36}\pi ^2
      -\frac{25}{3}\zeta (3)
      }{\epsilon }
    \\&\quad\quad\quad\quad\quad\quad\notag
    +\frac{5711}{162}
    +\frac{17}{9}\pi ^2
    -33 \zeta (3)
    -\frac{7}{60}\pi ^4
    \phantom{\Bigg\}}
    \\&\quad\quad\quad\quad\quad\quad\notag
    +\epsilon\Bigg(
       \frac{189767}{972}
      +\frac{65}{27}\pi ^2
      -\frac{1058}{27}\zeta (3)
      -\frac{1111}{2160}\pi ^4
      +\frac{71}{5}\zeta (5)
      +\frac{23}{18}\pi ^2 \zeta (3)
      \Bigg)
    \\&\quad\quad\quad\quad\quad\quad\notag
    +\epsilon^2\Bigg(
       \frac{4972715}{5832}
      -\frac{233}{324}\pi ^2
      -\frac{26404}{81}\zeta (3)
      -\frac{307}{360}\pi ^4
      -\frac{341}{5}\zeta (5)
      \\&\quad\quad\quad\quad\quad\quad\quad\quad\quad\notag
      +\frac{257}{1680}\pi ^6
      -\frac{11}{54}\pi ^2 \zeta (3)
      +\frac{901}{9}\zeta (3)^2
      \Bigg)
    \Bigg]
    \\&\quad\ \,\notag
+C_A N_F \Bigg[
   -\frac{7}{3 \epsilon ^3}
   -\frac{13}{3 \epsilon ^2}
   +\frac{
      \frac{76}{27}
     +\frac{\pi ^2}{18}
     }{\epsilon}
   -\frac{916}{81}
   -\frac{5}{18}\pi^2
   -2 \zeta (3)
   \\&\quad\quad\quad\quad\quad\quad\notag
   +\epsilon\Bigg(
     -\frac{14603}{243}
     -\frac{8}{27}\pi ^2
     -\frac{604}{27}\zeta (3)
     -\frac{59}{1080}\pi ^4
     \Bigg)
   \\&\quad\quad\quad\quad\quad\quad\notag
   +\epsilon^2\Bigg(
     -\frac{366023}{1458}
     +\frac{127}{162}\pi ^2
     -\frac{4448}{81}\zeta (3)
     -\frac{257}{648}\pi ^4
     -\frac{98}{5}\zeta (5)
     +\frac{61}{27}\pi ^2 \zeta (3)
     \Bigg)
   \Bigg]
   \\&\quad\ \,\notag
+C_F N_F \Bigg[
    \frac{1}{\epsilon }
   -\frac{73}{6}
   +8 \zeta (3)
   +\epsilon\Bigg(
     -\frac{2045}{36}
     +\frac{7}{18}\pi ^2
     +\frac{92}{3}\zeta (3)
     +\frac{4}{27}\pi ^4
     \Bigg)
   \\&\quad\quad\quad\quad\quad\quad\notag
   +\epsilon ^2\Bigg(
     -\frac{53269}{216}
     +\frac{263}{108}\pi ^2
     +\frac{1232}{9}\zeta (3)
     +\frac{46}{81}\pi ^4
     +32 \zeta (5)
     -\frac{20}{9}\pi ^2 \zeta (3)
     \Bigg)
  \Bigg]
\\&\quad\quad\quad\quad
+\frac{4 N_F^2}{9 \epsilon ^2}
\Bigg\}
+\mathcal{O}(\epsilon^3),
\\[15pt] \notag
\bar{F}^{\text{1L}}_{\gtilde} = \ &
\Big(\frac{\alphas}{4\pi}\Big)\Bigg\{
 C_A \Bigg[
   -\frac{2}{\epsilon ^2}
   -\frac{4}{\epsilon }
   +2
   +\frac{\pi ^2}{6}
   +\epsilon\frac{14}{3}\zeta (3)
   +\epsilon ^2\frac{47}{720}\pi ^4
   \\&\quad\quad\quad\quad\quad\, \notag
   +\epsilon^3 \Big(
      \frac{62}{5}\zeta (5)
     -\frac{7}{18}\pi ^2 \zeta (3)
     \Big)
   +\epsilon^4 \Big(
      \frac{949}{60480}\pi ^6
     -\frac{49}{9}\zeta(3)^2
     \Big)
   \Bigg]
 \\&\quad\quad\quad
 +\frac{N_F}{\epsilon}
 \Bigg\}
 +\mathcal{O}(\epsilon^5),
\\[15pt] \notag
\bar{F}^{\text{2L}}_{\gtilde} = \ &
\Big(\frac{\alphas}{4\pi}\Big)^2\Bigg\{
C_A^2 \Bigg[
   \frac{2}{\epsilon ^4}
  +\frac{27}{2 \epsilon ^3}
  +\frac{
    \frac{64}{9}
    -\frac{\pi ^2}{6}
    }{\epsilon ^2}
  +\frac{
    -\frac{1211}{54}
    -\frac{\pi ^2}{4}
    -\frac{25}{3}\zeta (3)
    }{\epsilon }
  \\&\quad\quad\quad\quad\quad\notag
  +\frac{6052}{81}
  +\frac{263}{108}\pi ^2
  -\frac{323}{9}\zeta (3)
  -\frac{7}{60}\pi ^4
  \\&\quad\quad\quad\quad\quad\notag
  +\epsilon\Bigg(
     \frac{263363}{972}
    +\frac{1489}{324}\pi ^2
    -\frac{1655}{27}\zeta (3)
    -\frac{67}{120}\pi ^4
    +\frac{71}{5}\zeta (5)
    +\frac{23}{18}\pi ^2 \zeta (3)
    \Bigg)
  \\&\quad\quad\quad\quad\quad\notag
  +\epsilon ^2\Bigg(
     \frac{6457043}{5832}
    +\frac{6803}{972}\pi ^2
    -\frac{34459}{81}\zeta (3)
    -\frac{15221}{12960}\pi ^4
    -\frac{235}{3}\zeta (5)
    \\&\quad\quad\quad\quad\quad\quad\quad\quad\notag
    +\frac{257}{1680}\pi ^6
    -\frac{16}{27}\pi ^2 \zeta (3)
    +\frac{901}{9}\zeta (3)^2
    \Bigg)
  \Bigg]
  \\&\notag
+C_A N_F \Bigg[
  -\frac{3}{\epsilon^3}
  -\frac{52}{9 \epsilon^2}
  +\frac{151}{27 \epsilon }
  -\frac{1925}{162}
  -\frac{25}{54}\pi ^2
  -\frac{28}{9}\zeta (3)
  \\&\quad\quad\quad\quad\ \,\notag
  +\epsilon\Bigg(
    -\frac{10538}{243}
    -\frac{46}{81}\pi ^2
    -\frac{922}{27}\zeta (3)
    -\frac{61}{720}\pi ^4
    \Bigg)
  \\&\quad\quad\quad\quad\ \,\notag
  +\epsilon ^2\Bigg(
    -\frac{291065}{1458}
    +\frac{419}{486}\pi ^2
    -\frac{8678}{81}\zeta (3)
    -\frac{3971}{6480}\pi ^4
  \\&\quad\quad\quad\quad\quad\quad\quad\ \,\notag
    -\frac{382}{15}\zeta (5)
    +\frac{203}{54}\pi ^2 \zeta (3)
    \Bigg)
  \Bigg]
  \\&\notag
+C_F N_F \Bigg[
  -\frac{1}{\epsilon ^2}
  +\frac{1}{2 \epsilon }
  -41
  -\frac{\pi ^2}{3}
  +12 \zeta (3)
  \\&\quad\quad\quad\quad\ \ \notag
   +\epsilon\Bigg(
    -\frac{669}{4}
    -\frac{3}{2}\pi ^2
    +\frac{196}{3}\zeta (3)
    +\frac{2}{9}\pi ^4
    \Bigg)
  \\&\quad\quad\quad\quad\ \ \notag
  +\epsilon^2\Bigg(
    -\frac{4607}{8}
    -\frac{61}{12}\pi ^2
    +\frac{868}{3}\zeta (3)
    +\frac{67}{60}\pi ^4
    +48 \zeta (5)
    -\frac{10}{3}\pi ^2 \zeta (3)
    \Bigg)
  \Bigg]
  \\&
+\frac{N_F^2}{\epsilon ^2}
\Bigg\}
+\mathcal{O}(\epsilon^3).
\end{align}

\section{Conclusions}
\label{sec:conclusions}

We have computed the $H\to gg$ amplitudes
at the two-loop level in the \FDH\ and \DRED\ scheme and presented the
$\RS$ renormalized on-shell results up to the order $\epsilon^2$. In
\DRED, this involves two different amplitudes for $H\to \ghat\ghat$
and $H\to \gtilde\gtilde$  with external gluons/$\epsilon$-scalars.
The computation is motivated because it contains key elements which
constitute important building blocks for further computations, and
because it is essential for the complete understanding of the
infrared divergence structure of \FDH\ and \DRED\ amplitudes.

The renormalization procedure has been described in detail. It is less trivial
than in many QCD calculations in \CDR, since not only the strong
coupling needs to be renormalized but also evanescent couplings of the
$\epsilon$-scalar. The computation provides a further example of
the well-known fact that regardless of whether \FDH\ or \DRED\ is used,
the evanescent couplings have to be renormalized independently.

Further, the renormalization of the effective dimension-5 operators
involves mixing with new, $\epsilon$-scalar dependent operators.
A suitable basis of operators has been provided. One unavoidable fact is
that the extended operator space contains operators which are total
derivatives. As a result the required operator mixing renormalization
constants cannot be obtained in the same elegant way of
Ref.~\cite{Spiridonov:1984br} as in \CDR.
Instead, they had to be obtained from explicit one- and two-loop
off-shell calculations.

The results for the UV renormalized but infrared divergent form factors
can also be used to complete the study of the general infrared divergence
structure of two-loop amplitudes in \FDH\ and \DRED, begun in
Ref.~\cite{Gnendiger:2014nxa,Kilgore:2012tb}. From general principles it is known
that all infrared divergences can be expressed in terms of cusp and
parton anomalous dimensions. The results of the present paper allow
to extract the final missing two-loop anomalous dimension for
external $\epsilon$-scalars. This extraction, together with further
checks and results, will be presented in a forthcoming paper
\cite{IRstructure}, where the infrared structure will also be investigated by
a SCET approach.

\subsection*{Acknowledgments}

We are grateful to M.\ Steinhauser and W.\ Kilgore for 
useful discussions. We acknowledge financial support from the DFG
grant STO/876/3-1.
A. Visconti is supported by the Swiss National Science
Foundation (SNF) under contract 200021-144252.

\begin{appendix}
\section{Appendix}
\subsection{Projectors and form factors of gluons and $\epsilon$-scalars}
\label{sec:appendix_A}
According to its Lorentz structure the on-shell Green-function
$\Gamma_{H\Ahat^\mu\Ahat^\nu}^{\text{on-shell}}$ can be represented as
\begin{align}
\Gamma_{H\Ahat^\mu\Ahat^\nu}^{\text{on-shell}} =
a\,(p\cdot r)\,\ghat^{\mu\nu}+ b\,p^\nu r^\mu + c\,p^\mu r^\nu + d\,p^\mu p^\nu + e\,r^\mu r^\nu,
\end{align}
where the coefficients $a\dots e$ are momentum-dependent quantities,
and coefficient $a$ is the gluon form factor.
Due to QCD Ward-identities the relation $a=-b$ holds, see e.\,g. Ref.~\cite{Harlander:2000mg}.
Accordingly,  the on-shell Green-function $\Gamma_{H\Atilde^\mu\Atilde^\nu}^{\text{on-shell}}$
with external $\epsilon$-scalars can be represented as
\begin{align}
\Gamma_{H\Atilde^\mu\Atilde^\nu}^{\text{on-shell}} =
f\,(p\cdot r)\,\gtilde^{\mu\nu},
\end{align}
where we refer to $f$ as $\epsilon$-scalar form factor.
All coefficients of the covariant decomposition can be extracted with
appropriate projection operators that are given below.

In the off-shell case the UV divergence structure of
$\Gamma_{H\Ahat^\mu\Ahat^\nu}$
can be represented in a more specific way as
\begin{align}
\Gamma_{H\Ahat^\mu\Ahat^\nu}\Big|^{\text{off-shell}}_{\text{UV div.}} =
\left[A+A'\,\frac{p^2+r^2}{(p\cdot r)}\right](p\cdot r)\,\ghat^{\mu\nu}
+ B\,p^\nu r^\mu + C\,p^\mu r^\nu + D\,p^\mu p^\nu + E\,r^\mu r^\nu,
\end{align}
where the coefficients $A\dots E$ are now momentum-independent.
Since these divergences can be absorbed by counterterms corresponding
to operators $O_1$ and $O_4$ the relation $A=-B$ again holds,
see e.\,g. Feynman rules (\ref{frO1a}) and (\ref{frO4a}).
Due to this there are two possibilities of extracting coefficient $A$,
which corresponds to the desired renormalization constant
$\delta Z^{\text{2L}}_{\lambda}$:
The first one is to extract the coefficient of $(p\cdot r)\,\ghat^{\mu\nu}$
and neglect terms $\propto p^2, r^2$;
the second is to extract coefficient $-B$.
We checked explicitly that the relations $a=-b$ and $A=-B$
hold throughout the paper.

Again, the covariant decomposition with external $\epsilon$-scalars is much simpler and reads:
\begin{align}
\Gamma_{H\Atilde^\mu\Atilde^\nu}\Big|^{\text{off-shell}}_{\text{UV div.}} = 
\left[F+F'\,\frac{p^2+r^2}{(p\cdot r)}\right](p\cdot r)\,\gtilde^{\mu\nu}.
\end{align}
The desired coefficient for the computation of 
$\delta Z^{\text{2L}}_{\leps}$ is $F$. 
Accordingly, we extract the coefficient of $(p\cdot r)\,\gtilde^{\mu\nu}$ and neglect terms
$\propto p^2, r^2$.

The corresponding projection operators are:
\begin{subequations}
\begin{align}
\begin{split}
 P^{\mu\nu}_{g,(p\cdot r)\ghat^{\mu\nu}} &= \Big\{
   \ghat^{\mu\nu}\left[(p\cdot r)^2-p^2r^2\right]\\
   &\quad\quad-(p^\nu r^\mu+p^\mu r^\nu)(p\cdot r)\\
   &\quad\quad+p^\mu p^\nu r^2+r^\mu r^\nu p^2\Big\}
   \frac{1}{(D-2)(p\cdot r)\left[(p\cdot r)^2-p^2r^2\right]},
 \end{split}\\
 \begin{split}
 P^{\mu\nu}_{g,p^\nu r^\mu} &=\Big\{
   \ghat^{\mu\nu}\,(p\cdot r)\left[p^2 r^2-(p\cdot r)^2\right]\\
   &\quad\quad+p^\nu r^\mu\left[(p\cdot r)^2+p^2 r^2 (D-2)\right]
   +p^\mu r^\nu\,(p\cdot r)^2\,(D-1)\\
   &\quad\quad+(p^\mu p^\nu r^2+r^\mu r^\nu p^2)(p\cdot r)(1-D)\Big\}
   \frac{1}{(D-2)\left[(p\cdot r)^2-p^2r^2\right]^2},
 \end{split}\\
 P^{\mu\nu}_{\gtilde,(p\cdot r)\gtilde^{\mu\nu}} &=
 \frac{\gtilde^{\mu\nu}}{\Neps(p\cdot r)}.
\end{align}
\end{subequations}

\subsection{Feynman rules}
\label{sec:appendix_B}
In the following we give Feynman rules according to operators
$O_1, \Otilde_1, O_4$ and $\Otilde_4$
that are needed for the renormalization in the \FDH\ and \DRED\ scheme.
Feynman rules including four $\epsilon$-scalars
are not relevant in this paper and are not given explicitly.
\begin{align}
\intertext{$\bullet$ Feynman rules according to the Lagrangian term $\lambda H O_1$:}
 \scalebox{.6}{
 \begin{picture}(200,0)(0,50)
 \Vertex(35,50){2}
 \DashLine(0,50)(35,50){2}
 \Gluon(35,50)(100,100){5}{6}
 \Gluon(35,50)(100,0){5}{6}
 \LongArrow(80,96)(60,80)
 \LongArrow(80,6)(60,21)
 \Text(63, 100)[c]{\scalebox{1.67}{$k_2$}}
 \Text(63, 5  )[c]{\scalebox{1.67}{$k_1$}}
 \Text( 25, 70)[c]{\scalebox{1.67}{$O_1$}}
 \Text(-15, 50)[c]{\scalebox{1.67}{$H$}}
 \Text(115, -5)[c]{\scalebox{1.67}{$\Ahat^{\alpha}_{a}$}}
 \Text(115,105)[c]{\scalebox{1.67}{$\Ahat^{\beta }_{b}$}}
 \label{frO1a}
 \end{picture} }
 &=\quad i\lambda\,\Big[(k_1\cdot k_2)\,\ghat^{\,\alpha\beta}-k_1^{\,\beta}\,k_2^{\,\alpha}\Big]\,\delta^{ab}
 \phantom{\begin{aligned}\bigg|\\ \bigg|\\ \bigg|\end{aligned}}
 \\
 \scalebox{.6}{
 \begin{picture}(200,0)(0,50)
 \Vertex(35,50){2}
 \DashLine(0,50)(35,50){2}
 \Gluon(35,50)(100,100){5}{5}
 \Gluon(35,50)(100,50){5}{4}
 \Gluon(35,50)(100,0){5}{5}
 \LongArrow(80,96)(60,80)
 \LongArrow(80,6)(60,21)
 \LongArrow(80,60)(60,60)
 \Text( 63,100)[c]{\scalebox{1.67}{$k_3$}}
 \Text( 95, 65)[c]{\scalebox{1.67}{$k_2$}}
 \Text( 63,  5)[c]{\scalebox{1.67}{$k_1$}}
 \Text( 25, 70)[c]{\scalebox{1.67}{$O_1$}}
 \Text(-15,50 )[c]{\scalebox{1.67}{$H$}}
 \Text(115, -5)[c]{\scalebox{1.67}{$\Ahat^{\alpha}_{a}$}}
 \Text(115, 50)[c]{\scalebox{1.67}{$\Ahat^{\beta }_{b}$}}
 \Text(115,105)[c]{\scalebox{1.67}{$\Ahat^{\gamma}_{c}$}}
 \end{picture} }
 &=\quad-\lambda\,\gs f^{abc}\times\left[ 
 \begin{aligned}
 & \ghat^{\alpha\beta }\left(k_1-k_2\right)^{\gamma}  \\ &
 + \ghat^{\beta\gamma }\left(k_2-k_3\right)^{\alpha}  \\ &
 + \ghat^{\gamma\alpha}\left(k_3-k_1\right)^{\beta}
 \end{aligned}
 \right]
 \phantom{\begin{aligned}\bigg|\\ \bigg|\\ \bigg|\end{aligned}}
 \\
 \scalebox{.6}{
 \begin{picture}(200,0)(0,50)
 \Vertex(35,50){2}
 \DashLine(0,50)(35,50){2}
 \Gluon(35,50)(100,100){5}{5}
 \Gluon(35,50)(100, 66){5}{5}
 \Gluon(35,50)(100, 33){5}{5}
 \Gluon(35,50)(100,  0){5}{5}
 \Text( 25, 70)[c]{\scalebox{1.67}{$O_1$}}
 \Text(-15, 50)[c]{\scalebox{1.67}{$H$}}
 \Text(115, -6)[c]{\scalebox{1.67}{$\Ahat^{\alpha}_{a}$}}
 \Text(115, 31)[c]{\scalebox{1.67}{$\Ahat^{\beta }_{b}$}}
 \Text(115, 68)[c]{\scalebox{1.67}{$\Ahat^{\gamma}_{c}$}}
 \Text(115,105)[c]{\scalebox{1.67}{$\Ahat^{\delta}_{d}$}}
 \end{picture} }
 &=\quad -i\lambda\, \gs^2\times\left[\begin{aligned}
 & \ghat^{\alpha\beta }\ghat^{\gamma\delta}\left(f^{ace}f^{bde}+f^{ade}f^{bce}\right)\\
 &+\ghat^{\alpha\gamma}\ghat^{\beta\delta }\left(f^{abe}f^{cde}-f^{ade}f^{bce}\right)\\
 &-\ghat^{\alpha\delta}\ghat^{\beta\gamma }\left(f^{abe}f^{cde}+f^{ace}f^{bde}\right)
 \end{aligned}\right]
 \phantom{\begin{aligned}\bigg|\\ \bigg|\\ \bigg|\end{aligned}}
\intertext{$\bullet$ Feynman rules according to the Lagrangian term $\leps H \Otilde_1$}
 \scalebox{.6}{
 \begin{picture}(200,0)(0,50)
 \Vertex(35,50){2}
 \DashLine(0,50)(35,50){2}
 \DashLine(35,50)(100,100){4}
 \DashLine(100,0)(35,50){4}
 \LongArrow(80,96)(60,80)
 \LongArrow(80,6)(60,21)
 \Text(63, 100)[c]{\scalebox{1.67}{$k_2$}}
 \Text(63, 5  )[c]{\scalebox{1.67}{$k_1$}}
 \Text( 25, 70)[c]{\scalebox{1.67}{$\Otilde_1$}}
 \Text(-15,50 )[c]{\scalebox{1.67}{$H$}}
 \Text(115, -5)[c]{\scalebox{1.67}{$\Atilde^{\,\alpha}_{a}$}}
 \Text(115,105)[c]{\scalebox{1.67}{$\Atilde^{\,\beta }_{b}$}}
 \end{picture} }
 &=\quad i\lambdaeps\,\Big[(k_1\cdot k_2)\,\gtilde^{\,\alpha\beta}\Big]\,\delta^{ab}
 \phantom{\begin{aligned}\bigg|\\ \bigg|\\ \bigg|\end{aligned}}
 \\
 \scalebox{.6}{
 \begin{picture}(200,0)(0,50)
 \Vertex(35,50){2}
 \DashLine(0,50)(35,50){2}
 \Gluon(35,50)(100,100){5}{5}
 \DashLine(35,50)(100,50){4}
 \DashLine(100,0)(35,50){4}
 \LongArrow(80,96)(60,80)
 \LongArrow(80,6)(60,21)
 \LongArrow(80,60)(60,60)
 \Text( 63,100)[c]{\scalebox{1.67}{$k_3$}}
 \Text( 95, 65)[c]{\scalebox{1.67}{$k_2$}}
 \Text( 63,  5)[c]{\scalebox{1.67}{$k_1$}}
 \Text( 25, 70)[c]{\scalebox{1.67}{$\Otilde_1$}}
 \Text(-15,50 )[c]{\scalebox{1.67}{$H$}}
 \Text(115, -5)[c]{\scalebox{1.67}{$\Atilde^{\alpha}_{a}$}}
 \Text(115, 50)[c]{\scalebox{1.67}{$\Atilde^{\beta }_{b}$}}
 \Text(115,105)[c]{\scalebox{1.67}{$\Ahat^{\gamma}_{c}$}}
 \end{picture} }
 &=\quad -\lambdaeps\,\gs f^{abc}\,\gtilde^{\,\alpha\beta}\,(k_1-k_2)^{\gamma}
 \phantom{\begin{aligned}\bigg|\\ \bigg|\\ \bigg|\end{aligned}}
 \\
 \scalebox{.6}{
 \begin{picture}(200,0)(0,50)
 \Vertex(35,50){2}
 \DashLine(0,50)(35,50){2}
 \Gluon(35,50)(100,100){5}{5}
 \Gluon(35,50)(100, 66){5}{5}
 \DashLine(35,50)(100, 33){4}
 \DashLine(35,50)(100,  0){4}
 \Text( 25, 70)[c]{\scalebox{1.67}{$\Otilde_1$}}
 \Text(-15, 50)[c]{\scalebox{1.67}{$H$}}
 \Text(115, -6)[c]{\scalebox{1.67}{$\Atilde^{\alpha}_{a}$}}
 \Text(115, 31)[c]{\scalebox{1.67}{$\Atilde^{\beta }_{b}$}}
 \Text(115, 68)[c]{\scalebox{1.67}{$\Ahat^{\gamma}_{c}$}}
 \Text(115,105)[c]{\scalebox{1.67}{$\Ahat^{\delta}_{d}$}}
 \end{picture} }
 &=\quad -i\lambdaeps\,\gs^2\, \gtilde^{\,\alpha\beta}\ghat^{\gamma\delta}\left(f^{ace}f^{bde}+f^{ade}f^{bce}\right)
 \phantom{\begin{aligned}\bigg|\\ \bigg|\\ \bigg|\end{aligned}}
\intertext{$\bullet$  Feynman rules according to the Lagrangian term $H O_4$:}
  \scalebox{.6}{
 \begin{picture}(200,0)(0,50)
 \Vertex(35,50){2}
 \DashLine(0,50)(35,50){2}
 \Gluon(35,50)(100,100){5}{5}
 \Gluon(35,50)(100,0){5}{5}
 \LongArrow(80,96)(60,80)
 \LongArrow(80,6)(60,21)
 \Text(63, 100)[c]{\scalebox{1.67}{$k_2$}}
 \Text(63, 5  )[c]{\scalebox{1.67}{$k_1$}}
 \Text( 25, 70)[c]{\scalebox{1.67}{$O_4$}}
 \Text(-15, 50)[c]{\scalebox{1.67}{$H$}}
 \Text(115, -5)[c]{\scalebox{1.67}{$\Ahat^{\alpha}_{a}$}}
 \Text(115,105)[c]{\scalebox{1.67}{$\Ahat^{\beta }_{b}$}}
 \label{frO4a}
 \end{picture} }
 &=\quad -i\,\Big[\left(k_1^2 + k_2^2\right)\ghat^{\,\alpha\beta}
   -\left(k_1^{\,\alpha}\,k_1^{\,\beta}+k_2^{\,\alpha}\,k_2^{\,\beta}\right)\Big]\,\delta^{ab}
   \phantom{\begin{aligned}\bigg|\\ \bigg|\\ \bigg|\end{aligned}}
 \\
 \scalebox{.6}{
 \begin{picture}(200,0)(0,50)
 \Vertex(35,50){2}
 \DashLine(0,50)(35,50){2}
 \Gluon(35,50)(100,100){5}{5}
 \Gluon(35,50)(100,50){5}{5}
 \Gluon(35,50)(100,0){5}{5}
 \LongArrow(80,96)(60,80)
 \LongArrow(80,6)(60,21)
 \LongArrow(80,60)(60,60)
 \Text( 63,100)[c]{\scalebox{1.67}{$k_3$}}
 \Text( 95, 65)[c]{\scalebox{1.67}{$k_2$}}
 \Text( 63,  5)[c]{\scalebox{1.67}{$k_1$}}
 \Text( 25, 70)[c]{\scalebox{1.67}{$O_4$}}
 \Text(-15, 50)[c]{\scalebox{1.67}{$H$}}
 \Text(115, -5)[c]{\scalebox{1.67}{$\Ahat^{\alpha}_{a}$}}
 \Text(115, 50)[c]{\scalebox{1.67}{$\Ahat^{\beta }_{b}$}}
 \Text(115,105)[c]{\scalebox{1.67}{$\Ahat^{\gamma}_{c}$}}
 \end{picture} }
 &=\quad3\,\gs f^{abc}\times\left[ 
 \begin{aligned}
 & \ghat^{\alpha\beta }\left(k_1-k_2\right)^{\gamma}  \\ &
 + \ghat^{\beta\gamma }\left(k_2-k_3\right)^{\alpha}  \\ &
 + \ghat^{\gamma\alpha}\left(k_3-k_1\right)^{\beta}
 \end{aligned}
 \right]
 \phantom{\begin{aligned}\bigg|\\ \bigg|\\ \bigg|\end{aligned}}
 \\
 \scalebox{.6}{
 \begin{picture}(200,0)(0,50)
 \Vertex(35,50){2}
 \DashLine(0,50)(35,50){2}
 \Gluon(35,50)(100,100){5}{5}
 \DashLine(35,50)(100,50){4}
 \DashLine(100,0)(35,50){4}
 \LongArrow(80,6)(60,21)
 \LongArrow(80,60)(60,60)
 \Text( 95, 65)[c]{\scalebox{1.67}{$k_2$}}
 \Text( 63,  5)[c]{\scalebox{1.67}{$k_1$}}
 \Text( 25, 70)[c]{\scalebox{1.67}{$O_4$}}
 \Text(-15, 50)[c]{\scalebox{1.67}{$H$}}
 \Text(115, -5)[c]{\scalebox{1.67}{$\Atilde^{\alpha}_{a}$}}
 \Text(115, 50)[c]{\scalebox{1.67}{$\Atilde^{\beta }_{b}$}}
 \Text(115,105)[c]{\scalebox{1.67}{$\Ahat^{\gamma}_{c}$}}
 \end{picture} }
 &=\quad -\gs f^{abc}\,\gtilde^{\,\alpha\beta}\,(k_1-k_2)^{\gamma}
 \phantom{\begin{aligned}\bigg|\\ \bigg|\\ \bigg|\end{aligned}}
 \\
 \scalebox{.6}{
 \begin{picture}(200,0)(0,50)
 \Vertex(35,50){2}
 \DashLine(0,50)(35,50){2}
 \ArrowLine(35,50)(100,100)
 \ArrowLine(100,50)(35,50)
 \Gluon(35, 50)(100,0){5}{5}
 \Text( 25, 70)[c]{\scalebox{1.67}{$O_4$}}
 \Text(-15, 50)[c]{\scalebox{1.67}{$H$}}
 \Text(115, -5)[c]{\scalebox{1.67}{$\Atilde^{\alpha}_{a}$}}
 \Text(115, 50)[c]{\scalebox{1.67}{$q_j$}}
 \Text(115,105)[c]{\scalebox{1.67}{$\overline{q}_i$}}
 \end{picture} }
 &=\quad -i\gs\,\gammahat^{\alpha}\left(T^a\right)_{ij}
 \phantom{\begin{aligned}\bigg|\\ \bigg|\\ \bigg|\end{aligned}}
 \\
  \scalebox{.6}{
 \begin{picture}(200,0)(0,50)
 \Vertex(35,50){2}
 \DashLine(0,50)(35,50){2}
 \DashArrowLine(35,50)(100,100){2}
 \DashArrowLine(100,0)(35,50){2}
 \LongArrow(80,96)(60,80)
 \LongArrow(80,6)(60,21)
 \Text(63, 100)[c]{\scalebox{1.67}{$k_2$}}
 \Text(63, 5  )[c]{\scalebox{1.67}{$k_1$}}
 \Text( 25, 70)[c]{\scalebox{1.67}{$O_4$}}
 \Text(-15, 50)[c]{\scalebox{1.67}{$H$}}
 \Text(115, -5)[c]{\scalebox{1.67}{$c_a$}}
 \Text(115,105)[c]{\scalebox{1.67}{$\overline{c}_b$}}
 \end{picture} }
 &=\quad i\,(k_1\cdot k_2)\,\delta_{ab}
 \phantom{\begin{aligned}\bigg|\\ \bigg|\\ \bigg|\end{aligned}}
 \intertext{$\bullet$ Feynman rules according to the Lagrangian term $H \Otilde_4$:}
 \scalebox{.6}{
 \begin{picture}(200,0)(0,50)
 \Vertex(35,50){2}
 \DashLine(0,50)(35,50){2}
 \DashLine(35,50)(100,100){4}
 \DashLine(35,50)(100,0){4}
 \LongArrow(80,96)(60,80)
 \LongArrow(80,6)(60,21)
 \Text(63, 100)[c]{\scalebox{1.67}{$k_2$}}
 \Text(63, 5  )[c]{\scalebox{1.67}{$k_1$}}
 \Text( 25, 70)[c]{\scalebox{1.67}{$\Otilde_4$}}
 \Text(-15, 50)[c]{\scalebox{1.67}{$H$}}
 \Text(115, -5)[c]{\scalebox{1.67}{$\Ahat^{\alpha}_{a}$}}
 \Text(115,105)[c]{\scalebox{1.67}{$\Ahat^{\beta }_{b}$}}
 \end{picture} }
 &=\quad -i\left(k_1^2 + k_2^2\right)\gtilde^{\,\alpha\beta}\,\delta^{ab}
 \phantom{\begin{aligned}\bigg|\\ \bigg|\\ \bigg|\end{aligned}}
 \\
 \scalebox{.6}{
 \begin{picture}(200,0)(0,50)
 \Vertex(35,50){2}
 \DashLine(0,50)(35,50){2}
 \Gluon(35,50)(100,100){5}{5}
 \DashLine(35,50)(100,50){4}
 \DashLine(100,0)(35,50){4}
 \LongArrow(80,6)(60,21)
 \LongArrow(80,60)(60,60)
 \Text( 95, 65)[c]{\scalebox{1.67}{$k_2$}}
 \Text( 63,  5)[c]{\scalebox{1.67}{$k_1$}}
 \Text( 25, 70)[c]{\scalebox{1.67}{$\Otilde_4$}}
 \Text(-15, 50)[c]{\scalebox{1.67}{$H$}}
 \Text(115, -5)[c]{\scalebox{1.67}{$\Atilde^{\alpha}_{a}$}}
 \Text(115, 50)[c]{\scalebox{1.67}{$\Atilde^{\beta }_{b}$}}
 \Text(115,105)[c]{\scalebox{1.67}{$\Ahat^{\gamma}_{c}$}}
 \end{picture} }
 &=\quad -\,2\,\gs f^{abc}\,\gtilde^{\,\alpha\beta}\,(k_1-k_2)^{\gamma}
 \phantom{\begin{aligned}\bigg|\\ \bigg|\\ \bigg|\end{aligned}}
\end{align}

\end{appendix}

\bibliography{bibliography}{}
\bibliographystyle{JHEP}

 \end{document}